\documentclass[aps,prb,notitlepage,amsfonts,citeautoscript,superscriptaddress,showpacs,twocolumn]{revtex4-1}

\usepackage{amsmath,amssymb,mathrsfs}
\usepackage{latexsym}
\usepackage{graphicx} 
\usepackage{epstopdf}
\usepackage{graphicx,epstopdf,color}
\usepackage{amsfonts}
\usepackage{amsmath,amssymb,mathrsfs}
\usepackage{hyperref}
\usepackage{caption}
\usepackage{color}
\usepackage{bm}

\newcommand{\nn}{\nonumber}

\usepackage{xr}

\begin{document}

\newcommand{\MIMe}{\textit{Rung Mott -- Meissner }}
\newcommand{\MI }{\textit{Rung Mott }}
\newcommand{\Me}{\textit{Meissner }}
\newcommand{\SF}{\textit{Rung superfluid }}
\newcommand{\VL}{\textit{Vortex lattice }}
\newcommand{\Laughlin}{\textit{Laughlin }}
\newcommand{\SFVL}{\textit{Rung superfluid -- Vortex lattice }}
\newcommand{\SFMe}{\textit{Rung superfluid -- Meissner }}
\newcommand{\MIVL}{\textit{Rung Mott -- Vortex lattice }}
\newcommand{\SMe}{\textit{Spinon Meissner }}
\newcommand{\SC}{\textit{SC proximity effect }}
\newcommand{\MISMe}{\textit{Rung Mott -- Spinon Meissner }}

\title{Chiral Mott Insulators, Meissner Effect, and Laughlin States in Quantum Ladders}

\author{Alexandru Petrescu}
\affiliation{Department of Physics, Yale University, New Haven, CT 06520, USA}
\affiliation{Centre de Physique Th\' eorique, \' Ecole Polytechnique, CNRS, 91128 Palaiseau C\' edex, France}

\author{Karyn Le Hur}
\affiliation{Centre de Physique Th\' eorique, \' Ecole Polytechnique, CNRS, 91128 Palaiseau C\' edex, France}

\begin{abstract}
We introduce generic bosonic models exemplifying that chiral Meissner currents can persist in insulating phases of matter. We first consider interacting bosons on a two-leg ladder. The total density sector can be gapped in a bosonic Mott insulator at odd-integer filling, while the relative density sector remains superfluid due to interchain hopping. Coupling the relative density to gauge fields yields a pseudospin Meissner effect. We show that the same phase arises if the bosons are replaced by spinful fermions confined in Cooper pairs, and find a dual fermionic Mott insulator with spinon currents. We prove that by tuning the mean density the Mott insulator with Meissner currents turns into a low-dimensional bosonic $\nu = \frac{1}{2}$ Laughlin state for strong enough repulsive interactions across the ladder rungs. We finally discuss extensions to multileg ladders and bilayers in which spinon superfluids with Meissner currents become possible. We propose two experimental realizations, one with ultracold atoms in the setup of Atala \textit{et al.}, Nat.~Phys.~\textbf{8}, 588 (2014) and another with Josephson junction arrays. We also address a Bose-Fermi mixture subject to a magnetic field in connection with the pseudo-gap phase of high-Tc cuprates.
\end{abstract}
\date{\today}
\maketitle


\section{Introduction}
On a lattice at commensurate filling, a bosonic Mott insulator \cite{Fisher1989, Greiner2001} is a state that can be adiabatically connected to an atomic insulator. In the Hamiltonian describing the atomic limit the kinetic terms providing tunneling between distinct sites are suppressed; consequently, the ground state is a product Fock state in which the variance of particle number at each site vanishes. Recent studies examine the possibility of nontrivial Mott states which, due to a broken symmetry, exhibit chiral current order and therefore quantum entanglement. Chiral Mott insulators have been shown to be closely related to short-range entangled topological phases of bosons, such as the boson topological insulator \cite{XuSenthil2013,*LiuGuWen2014}. The boson topological insulator is a symmetry protected topological phase \cite{Haldane1983, *AKLT1987, LuVishwanath,*Chen2013,*SenthilReview,*RegnaultSenthil2013}, whose gapless boundary excitations are protected by bulk symmetry, but do not posses topological order.

A route in the quest for nontrivial Mott insulators is to break time-reversal symmetry manifestly by an external magnetic field. With two bosonic species on the lattice, the external field may be coupled to the pseudospin degree of freedom within a Mott phase of total density \cite{Cole2012,*Radic2012,*Mandal2012,*Cai2012}. The Mott phase of spinful fermions in the time-reversal invariant Hofstadter model with additional Rashba spin-orbit coupling possesses spiral spin order \cite{Cocks2012,*Orth2013}. The unit-filled bosonic Haldane model \cite{Haldane1988} sustains a Mott insulator with nontrivial plaquette currents \cite{Vasic2014}.  In our previous work \cite{1}, we exemplify that the Josephson effect in the pseudospin sector leads to extended Meissner currents or a vortex lattice \cite{OrignacGiamarchi, Kardar1986}, while the total density retains Mott insulator correlations. This result was recently confirmed and extended numerically \cite{Piraud2014}. The Meissner effect has been recently probed experimentally \cite{Atala} with $^{87}$Rb atoms on a ladder optical lattice. 

\begin{figure}[t!]
\includegraphics[width=0.90\linewidth]{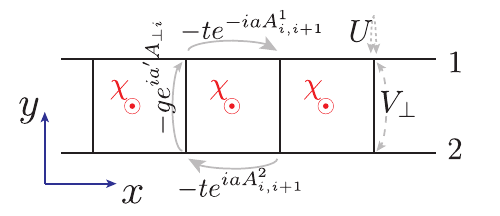}
\caption{\label{Fig:Ladder}
Energy scales and setup of bosonic two-leg ladder. Flux $\chi$ threads each square plaquette. Hopping integrals (\textit{solid arrows}) $t, g$ and repulsive interaction strengths (\textit{dashed arrows}) $U, V_\perp$ correspond to Eq.~(\ref{Eq:Hbt}). }
\end{figure}

In the presence of kinetic frustration, the superfluid state can spontaneously break a symmetry since the condensate forms over a linear combination of degenerate minima in the single particle spectrum. More importantly, for strong interactions, the broken symmetry may not be restored. This leads to Mott insulators with a spontaneously broken discrete symmetry, such as time-reversal. Such chiral Mott insulators have been predicted in quasi-one dimensional systems \cite{Dhar2012,*Dhar2013, Zaletel2014, TokunoGeorges}, and in two dimensions \cite{Aji2007,*Chakravarty2002,*Chua2011,*Messio2012,*Yan2011,*Huerga2013}

Interest in time-reversal symmetry breaking phases has been fueled by recent progress in the realization of artificial gauge fields in ultracold atoms \cite{Dalibard2011, Goldman2013} and photonic systems \cite{LuReview2014}. The quest for lattice equivalents of integer quantum Hall phases (with \cite{Hofstadter} or without \cite{Haldane1988} Landau levels) has led to implementations of artificial gauge fields with ultracold atoms \cite{AidelsburgerHofstadter,*Aidelsburger2014,*Miyake2013,*Jotzu,*JimenezGarcia1D,*Struck2012}, gyromagnetic photonic crystals at microwave frequency \cite{RaghuHaldane2008,*Wang2009,*LuReview2014}, coupled resonator optical waveguides \cite{Hafezi2011,*Hafezi2013,*Mittal2014}, metamaterials based on pillar-shaped photonic waveguides \cite{Khanikaev2012,*Rechtsman2013,*Sala2014}, optomechanical systems \cite{Schmidt2013}, or radio frequency devices \cite{Jia2013,*Albert2014}. Similar topological phases have been theoretically predicted to appear in Circuit Quantum Electrodynamics \cite{Koch2010,*Petrescu2012}. In tunable systems such as these band topology and edge transport can be probed. The interplay of a strong magnetic field and filling leads to the fractional quantum Hall effect, or the closely related spin liquids \cite{Balents2010}. Originally discovered in two-dimensional electron gases \cite{TSG1982}, the fractional quantum Hall effect has eluded implementation in quantum emulators, despite multiple theoretical proposals suitable for ultracold atoms \cite{Sorensen,*Hafezi,*Hormozi,*PalmerJaksch,*CooperDalibard2013,*Yao2013,*Liu2013,*Sterdyniak2014} or photons \cite{Hafezi2013,*Kapit2014}.

Motivated by the recent realization of the low-dimensional Meissner effect with ultracold atoms \cite{Atala}, we study an interacting boson tight-binding model near half filling and at arbitrary flux on a two-leg ladder (depicted in Fig.~\ref{Fig:Ladder}). This model is a generalization of our study in \cite{1}. There we showed that a half filled bosonic ladder with repulsive Hubbard interactions stabilizes a Mott insulating phase for total charge (+ sector), but which allows charge neutral Meissner currents in the relative density (- sector). For brevity, we will denote this phase by \MIMe. In this work, we extend our previous result in a number of ways. We first show that given certain commensuration conditions and in the presence of nearest neighbor repulsive interactions, the ground state corresponds to a coupled wire realization of the \Laughlin state \cite{Laughlin} introduced by Kane \textit{et al.} \cite{Kane2002}.  Analogous phases are supported in spinful fermion ladders, and a duality transformation allows us to determine a distinct class of spin chiral incompressible phases for fermions. While in the beginning we focus on quantum ladders, we construct analogous phases in two-dimensional lattices. We propose two feasible experimental setups in quantum circuits and in ultracold atoms. 

\begin{figure}[t!]
\includegraphics[width=1.00\linewidth]{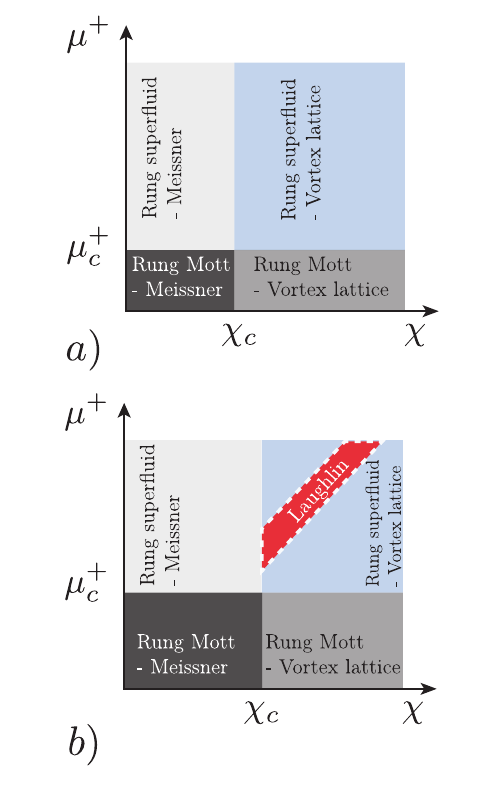}
\caption{\label{Fig:PDWB}Schematic phase diagram of the Josephson ladder: a) with $V_\perp = 0$ and no other repulsive long range interaction along the chains and b) for sufficiently strong $V_\perp \neq 0$ (or for long ranged repulsion within chains), it is possible to stabilize the \Laughlin ground state. The red region corresponds to~(\ref{Eq:LaCs}) with the lower sign. As $V_\perp$ is increased from $0$, the Mott gap and the critical flux $\chi_c$ increase. The phase notations are defined in Tables~\ref{Tab:PJL} and~\ref{Tab:PJLGaps}.}
\end{figure}

Our paper is organized as follows. In Sec.~\ref{Sec:WB} we derive the phase diagram in Figure~\ref{Fig:PDWB} for a bosonic ladder at or near half-odd integer filling per site.  Then, Sec.~\ref{Sec:Obs} contains discussions of observables, such as currents and flux quantization, which distinguish the possible ground states.  Next, we extend our results to spinful fermion ladders (potentially related to high-Tc cuprates) in Sec.~\ref{Sec:WF}. We propose in Sec.~\ref{Sec:Expt} two experimental realizations in Josephson junction arrays and ultracold atoms in optical lattices.  Sec.~\ref{Sec:Nleg} generalizes the phases found for two-leg ladders to $N$-leg ladders and bilayers. We summarize our results in the concluding Sec.~\ref{Sec:Conc}. Technical details in the Appendices will be referred to when necessary.


\section{Chiral phases of the Josephson ladder}
\label{Sec:WB}

\begin{table}
\begin{tabular}{|l|l|l|}
\hline
Sector & Notation & sine-Gordon term        \\
\hline
+      & \MI   &   $\sqrt{8}\phi^+$ \\
+      & \SF   &   $\sqrt{2}\phi^+ + 2\pi \delta n x$    \\
-      & \Me   &   $\sqrt{2}\theta^-$     \\
-      & \VL   &   $\sqrt{2}\theta^- + \chi x$    \\
+ \& - & \Laughlin      &   $\sqrt{2}\theta^- - \sqrt{8} \phi^+$       \\
\hline
\end{tabular}
\caption{\label{Tab:PJL} Phases of the Josephson ladder 
appearing in the phase diagram of Fig.~\ref{Fig:PDWB}. The
``+/-'' sector denotes total/relative vertical bond (rung) density $n_i^1 +/- n_i^2$ 
(see Fig.~\ref{Fig:Ladder}). The ``+'' sector can be in a Mott insulator
or superfluid phase, whereas the ``-'' sector can be in a Meissner phase or 
a vortex lattice phase depending on the strength of the field. The Laughlin
phase arises from a condition that mixes the two sectors. } 
\end{table}

\begin{table}
\begin{tabular}{|l|l|l|}
\hline
Phase & Gapped modes per & Gapless  \\
&  Sector      & modes    \\
\hline
\MIMe & $1^+ 1^-$   & $0$ \\
\MIVL & $1^+ 0^-$   & $1$ \\
\SFMe & $0^+ 1^-$   & $1$ \\
\SFVL & $0^+ 0^-$   & $2$ \\
\Laughlin        & $1^{+ \& -}$& $1$ \\
\hline
\end{tabular}
\caption{\label{Tab:PJLGaps} Number of gapped modes in the Josephson
ladder, for the phases appearing in Fig.~\ref{Fig:PDWB}. The only gapped
phase is \MIMe, the Mott insulator with Meissner currents. } 
\end{table}

In this section, we introduce a relatively simple insulating system exhibiting the Meissner effect \cite{MeissnerOchsenfeld,*Anderson1958,*deGennes}. We consider a bosonic quantum ladder with an odd number of bosons per rung (Fig. \ref{Fig:Ladder}). The two-leg ladder consists of two one-dimensional chains with inter- and intrachain kinetic and interaction terms. The lattice layout is depicted in Figure~\ref{Fig:Ladder}, which also summarizes the terms in the Hamiltonian   
\begin{eqnarray}
\label{Eq:Hbt}
H &=& -t \sum_{\alpha, i}  e^{i a A^\alpha_{i,i+1}}b_{\alpha i}^\dagger b_{\alpha , i+1}  - g \sum_{i} e^{-i a' A_{\perp i}} b^\dagger_{2 i} b_{1 i} + h.c.,  \nonumber \\
&+& \frac{U}{2} \sum_{\alpha,i} n_{\alpha i} (n_{\alpha i} - 1) + V_\perp \sum_i n_{1 i}n_{2 i} - \mu \sum_{\alpha i} n_{\alpha i}.
\end{eqnarray}
In Eq.~(\ref{Eq:Hbt}), the operator $b_{\alpha i}^\dagger$ creates a boson at site $i$ in chain $\alpha = 1,2$. We introduced the Peierls phases $a A^\alpha_{i,i+1}$ acquired by a particle on chain $\alpha=1,2$, and $a' A_{\perp i}$ between chains. Lengths $a$ and $a'$ are lattice spacings along and between chains; see Fig.~\ref{Fig:Ladder}. The spatial indices run between $1,..., L$, with $L+1 \equiv 1$ for periodic boundary conditions. $U$ and $V_\perp$ are repulsive on-site and ``rung'' interactions. 

The Hamiltonian~(\ref{Eq:Hbt}) without interactions can be realized in photonic systems (for a review see \cite{LuReview2014}). The weakly interacting limit has been realized experimentally in an ultracold atom ladder \cite{Atala}. Another possibility is to realize~(\ref{Eq:Hbt}) in Josephson junction arrays \cite{FazioVanDerZant2001} or more generally quantum circuits \cite{Devoret, DevoretMartinis}.

We briefly outline other results on the model in Eq.~(\ref{Eq:Hbt}) complementing our own \cite{1}. Without gauge fields and if $V_\perp = 0$, and $\mu$ insuring one boson per site, the model transitions from a Mott insulator to a superfluid as $g$ increases \cite{DonohueGiamarchi}.  At arbitrary boson filling and uniform flux there is a transition from the low-field Meissner phase to a vortex phase \cite{OrignacGiamarchi} beyond some critical field strength, reminiscent of type-II superconductivity. The low-field model with $V_\perp=0$ at unit filling exhibits a superfluid with Meissner currents and a Mott insulator with Meissner currents for weak enough $U$ \cite{TokunoGeorges}. The ground state for half a flux quantum per plaquette at integer filling is a chiral superfluid, a chiral Mott insulator or a Mott insulator, as argued by Dhar \textit{et al.} \cite{Dhar2012,*Dhar2013} and Tokuno and Georges \cite{TokunoGeorges}. The same model in the weakly interacting limit supports a staggered pattern of quantized orbital current vortices \cite{Lim2008,*Lim2010}.  The model without flux and $V_\perp = 0$ was studied for bosons with a hardcore constraint on site, $U \to \infty$ \cite{Crepin}. The ground state was shown to be a rung Mott insulator at half-filling. A recent numerical investigation covers the phase diagram versus filling, flux and interchain tunneling, containing a Meissner Mott insulator and a vortex lattice Mott insulator at half-filling \cite{Piraud2014}.

In this section we will uncover the ground state of the model~(\ref{Eq:Hbt}) at odd boson filling per rung, i.e. $2N+1$ bosons every two sites for $N \geq 0$ an integer. For $V_\perp = 0$, depending on different values of filling, flux, and interactions, we find the following low field phases: \MIMe, \SFMe, \MIVL, \SFVL.

While it is unnecessary for the phases listed above, the repulsive interaction $V_\perp > 0$ [or long-ranged repulsion within chain $\alpha$, not listed in Eq.~(\ref{Eq:Hbt})] changes slightly the phase diagram in that it controls the size of the gap above \MIMe. In the limit of large interactions, one can draw an analogy with a spin Meissner effect \cite{1}. Moreover, for large enough $V_\perp$, the ground state turns into a low-dimensional \Laughlin state if flux and doping are commensurate.  We note that a Hamiltonian related to (\ref{Eq:Hbt}) whose ground state is well approximated by the Laughlin state at $\nu=\frac{1}{2}$ has been discussed by Kalmeyer and Laughlin \cite{KalmeyerLaughlin1987,*KalmeyerLaughlin1989}, in search for a spin liquid ground state for the frustrated Heisenberg antiferromagnet. This theory was developed in succeding work including the formulation of a Hamiltonian whose exact ground state is the Laughlin state at $\nu=\frac{1}{2}$. \cite{Laughlin1989,*Schroeter2007} In this work we will identify the Laughlin state by comparing the continuum form of our Hamiltonian with that of a coupled wire construction \cite{Kane2002,TeoKane}.

The remainder of this section is structured as follows. Subsec.~\ref{Subsec:GInv} contains a discussion of the continuum limit and gauge invariance. Subsec.~\ref{Subsec:BMM} discusses the \Me phase. In Subsec.~\ref{Subsec:MI} we address the \MI transition within the Meissner state. Subsec.~\ref{Subsec:StabilityMIMe} addresses the stability of the \MIMe phase. In Subsec.~\ref{Subsec:La2} we introduce the condition that favors a low dimensional form of the \Laughlin state.

\subsection{Continuum limit and gauge invariance}
\label{Subsec:GInv}
In what follows we will derive a continuum, or bosonized form \cite{Giamarchi, Haldane1981, *GogolinNersesyanTsvelik} of Eq.~(\ref{Eq:Hbt}). We will be using the conventions of Ref.~\cite{Giamarchi} throughout this paper. The resulting field theory will allow us to treat interactions nonperturbatively and determine the possible ground states of the model in Eq.~(\ref{Eq:Hbt}). We begin by expressing the boson annihilation operator as $\psi^\alpha(x) = b_{\alpha j} / \sqrt{a}$, when $x = j a$. Then the bosonic creation operator in chain $\alpha$ becomes, in terms of new bosonic fields $\theta$ and $\phi$
\begin{equation}
\label{Eq:BosonOp}
(\psi^\alpha)^\dagger(x) = \sqrt{n_0^\alpha}  \sum_p e^{ i 2 p ( n^\alpha_0 \pi - \phi^\alpha)} e^{- i \theta^\alpha(x) }. 
\end{equation} 
We sum over all integers $p$. The field $\theta^\alpha(x)$ is the phase of the boson operator, whereas $\phi^\alpha(x')$ describes deviations from the mean density: $\delta n^\alpha \equiv n^\alpha - n_{0}^\alpha = -\frac{1}{\pi} \nabla \phi^\alpha$. The mean densities $n_0^{1,2}$ should be taken equal in practice, $n_0^{1,2} = n_0$. However, we shall keep the dependence on $n_0^{1,2}$ explicit to obtain more general expressions. The doublet $\theta,\phi$ satisfies the algebra
\begin{equation}
\label{Eq:A}
\left[ \phi^\alpha ( x ), \theta^\beta( x' )  \right] = i \frac{\pi}{2} \delta_{\alpha\beta} \text{Sign}( x' - x ).
\end{equation} 
For the ladder Hamiltonian it is convenient to introduce rotated fields
\begin{equation}
\label{Eq:Rotation}
\theta^\pm = (\theta^1 \pm \theta^2)/\sqrt{2} , \; \phi^\pm = (\phi^1 \pm \phi^2)/\sqrt{2}.
\end{equation}
These obey the same algebra in~(\ref{Eq:A}) for the new indices $\alpha = \pm$. In this basis, the model~(\ref{Eq:Hbt}) becomes
\begin{equation}
\label{Eq:HWB}
\mathcal{H} = \mathcal{H}_0^+ + \mathcal{H}_0^- + \mathcal{H}_{SG}.
\end{equation}
The first and second terms are Luttinger liquid Hamiltonians \cite{Giamarchi} for the symmetric and antisymmetric sectors, respectively:
\begin{eqnarray}
\label{Eq:LLp}
\mathcal{H}_0^+ = \frac{v^+}{2\pi} \int dx \left[ K^+ (\nabla \theta^+)^2 + \frac{1}{K^+} (\nabla \phi^+)^2  \right], \\
\label{Eq:LLm}
\mathcal{H}_0^- = \frac{v^-}{2\pi} \int dx \left[ K^- (\nabla \theta^- + A_\parallel^- )^2 + \frac{1}{K^-} (\nabla \phi^-)^2  \right].
\end{eqnarray}
$A_\parallel^-$ is a gauge field component whose line integral yields $A^\alpha_{ij}$. It will be discussed shortly. Note that in our geometry the artificial gauge field only couples to the antisymmetric (pseudo spin) sector. Under rotation~(\ref{Eq:Rotation}), and for $0 < V_\perp < U$, velocities of excitations and Luttinger parameters are expressed as:
\begin{eqnarray}
\label{Eq:vKWB}
v^\pm &=& v (1\pm V_\perp/U)^{1/2}, \nonumber \\
K^\pm &=& K (1\pm V_\perp/U)^{-1/2}.
\end{eqnarray}
In fact, a weak coupling Gross-Pitaevskii approximation of the bosonic operators in Eq.~(\ref{Eq:BosonOp}), followed by a gradient expansion, would allow us to identify $v = a\sqrt{tU}$ and $K=\sqrt{t/U}$ when $n_0^\alpha = \frac{1}{2a}$.  However, for general microscopic parameters in~(\ref{Eq:Hbt}), the Luttinger parameter satisfies $1 < K$ for repulsive interactions, $K = \infty$ for free bosons, $K<1$ for repulsive long-range interactions, and $K = 1$ for the hard core limit \cite{Giamarchi}.

The third term of Eq.~(\ref{Eq:HWB}) is a sine-Gordon (Josephson) Hamiltonian arising from the interchain coupling. Denoting the gauge field component in the $y$ direction by $A_\perp$, the coupling Hamiltonian reads 
\begin{eqnarray}
\label{Eq:HSGAparAperp}
\mathcal{H}_{SG} = - 2 g \sqrt{n_0^1 n_0^2} \int dx  \cos( - \sqrt{2} \theta^-  + a' A_\perp ) \times  \\
\left[ 1 + 2\cos\left( 2\pi  n_0^1 x - 2 \phi^1 \right)  \right] \, \left[ 1 + 2\cos\left( 2\pi  n_0^2 x - 2 \phi^2 \right)  \right].  \nonumber
\end{eqnarray}
$n_0^{\alpha}$ represent the mean density in each chain. The values of $A_\perp$, $A_\parallel$, $n_0^\alpha$ determine which contributions are to be considered from $\mathcal{H}_{\textit{SG}}$, based on lattice commensuration conditions.

We used in Eq.~(\ref{Eq:LLm}) the antisymmetric combination of gauge fields 
\begin{equation}
A_\parallel^- =  \frac{A^1_\parallel - A^2_\parallel}{\sqrt{2}}. 
\end{equation}
By convention, we require that the field $A_\parallel^\alpha(x)$ is related to the lattice gauge field $A_{ij}^{\alpha}$ of Eq.~(\ref{Eq:Hbt}) by an average over a straight line path between sites $j$ and $j+1$ on chain $\alpha$:
\begin{equation}
\int_{ja}^{(j+1)a} dx A^\alpha_\parallel(x) = a A_{j,j+1}^\alpha.
\end{equation}
Similarly, the component $A_\perp(x)$ appearing in Eq.~(\ref{Eq:HSGAparAperp}) is related to $A_{\perp,i}$ of Eq.~(\ref{Eq:Hbt}):
\begin{equation}
\int_\text{rung at $i$} dy A_\perp(y) = a' A_{\perp i}.
\end{equation}
The integral is performed over a rung at position $i$, starting from chain 1 and ending on chain 2. 

Ground state expectation values will only depend on the curl of the gauge field
\begin{equation}
\text{curl}
A = \nabla A_\perp(x) - \frac{A_\parallel^2(x) - A_\parallel^1(x)}{a'}.
\end{equation}
The lattice curl defines the flux through the plaquette
\begin{equation}
\text{curl}
A 
= \frac{\chi}{a'}.
\end{equation}
This equality defines the uniform flux perpendicular to the plane of the ladder. The plaquette enclosed between the rungs $j$ and $j+1$ is threaded by flux $a \chi = a a' \text{curl} A$. The Hamiltonian~(\ref{Eq:HWB}) is invariant under the gauge transformation
\begin{eqnarray}
\label{Eq:AparGT}
\tilde{A}^\alpha_\parallel(x) &=& A^\alpha_\parallel(x) + \nabla f^\alpha (x), \;   \\
\label{Eq:AperpGT}
\tilde{A}_\perp(x) &=& A_\perp(x) + \frac{f^2(x) - f^1(x)}{a'},  \\ 
\label{Eq:ThetaGT}
\tilde{\theta}^\alpha(x) &=& \theta^\alpha(x) - f^\alpha(x).
\end{eqnarray}
This preserves the algebra in Eq.~(\ref{Eq:A}). 

In the following treatment, it is favorable to use the gauge 
\begin{equation}
a' A_\perp(x) = \chi x, \;\; A_\parallel^\alpha = 0.
\end{equation}
The Hamiltonian in Eq.~(\ref{Eq:HSGAparAperp}) becomes 
\begin{eqnarray}
\label{Eq:HSG}
&&\mathcal{H}_{SG} = - 2 g \sqrt{n_0^1 n_0^2} \int dx  \cos( - \sqrt{2} \theta^-  + \chi x ) \times  \\
&&\left[ 1 + 2\cos\left( 2\pi  n_0^1 x - 2 \phi^1 \right)  \right] \, \left[ 1 + 2\cos\left( 2\pi  n_0^2 x - 2 \phi^2 \right)  \right].  \nonumber
\end{eqnarray}

We summarize the notations of the phases allowed by Eq.~(\ref{Eq:HSG}) in Tables~\ref{Tab:PJL} and \ref{Tab:PJLGaps}. A detailed discussion follows, but we anticipate the possible ground states here (see Figure~\ref{Fig:PDWB} for phase diagrams): 
At infinitesimal fluxes $\chi$, the cosine $\theta^-$ establishes Josephson phase coherence between the chains (\Me phase). When the flux per plaquette is high, the phases follow the variations of the gauge field, giving way to a \VL phase. Turning to the charge sector, at filling factors satisfying $n_0^1 + n_0^2 = \frac{2N+1}{a}$, where $N$ is nonnegative and integer, the cosine potential in $\phi^+$ favors an insulating ground state for total rung density, denoted \MI. At incommensurate fillings, this turns into a \SF.  We will introduce another state which exists if repulsive long ranged interactions are present. This state corresponds to a combined pinning of phase and charge fluctuations. We will denote it \Laughlin since it arises from a coupled wire construction \cite{Kane2002} of the bosonic Laughlin state at $\nu = 1/2$ \cite{Laughlin}. 

\subsection{\Me phase}
\label{Subsec:BMM}
The description of the phase diagram follows with the application of a two step renormalization group procedure. The renormalization group equations for Eq.~(\ref{Eq:HSG}) are solved in more detail in \cite{Crepin2011}. Here we provide an approximate solution which captures the essential physics.

First, we follow Ref. \cite{OrignacGiamarchi} and we focus on the term in Eq.~(\ref{Eq:HSG}) which is the most relevant in the renormalization group sense. This is a Josephson phase pinning between the two condensate phases $\theta^1$ and $\theta^2$:
\begin{equation}
\label{Eq:JosephsonSG}
\mathcal{H}_{SG} = -2 g \sqrt{n_0^1 n_0^2} \int dx \; \cos( - \sqrt{2}\theta^- + \chi x ).
\end{equation}

The renormalization group treatment to second order in the coupling $g$ is detailed in Appendix~\ref{Ap:RGE}.  We assume that $\chi a \ll 1$ such that the oscillatory argument in Eq.~(\ref{Eq:JosephsonSG}) is negligible. We define the dimensionless coupling constant (in units of the bandwidth) $g^- \equiv \frac{ga}{v}$. It flows to strong coupling if its bare value is nonzero and if its scaling dimension $1/(2 K^-)$ is less than 2. Assuming small temperatures $T \to 0$, the renormalization of the coupling constant $g^-$ is stopped at energy scales equal to the gap associated with the Josephson phase pinning. Inverting the RG equation for $g^-$, the gap has the following expression
\begin{equation}
\label{Eq:Deltam}
\Delta^- \sim \frac{v}{a} \left( g^- \right)^{\frac{1}{2-\frac{1}{2 K^-}}}.
\end{equation}
Here, we have approximated that $K^-$ renormalizes insignificantly. Therefore Eq.~(\ref{Eq:Deltam}) contains the bare coupling constant and Luttinger parameter. 

For temperatures $T<\Delta^-$, the field $\theta^-(x)$ is pinned to its classical value $\langle \theta^-(x)  \rangle =  \chi x $, leading to a vanishing of the interchain current and a saturation of intrachain currents \cite{OrignacGiamarchi}. Eq.~(\ref{Eq:Deltam}) implies that Josephson phase coherence between the chains occurs as soon as a nonzero tunneling matrix element $g$ is turned on; moreover, the gap above this ground state is a power law in the bare coupling $g$.

We denote this state by \Me. This phase is associated with gapped excitations of the external gauge field \cite{Anderson1963}. We illustrate this in the present situation considering the action for $\theta^-$. This is obtained easily from the Hamiltonian~(\ref{Eq:LLm}) and~(\ref{Eq:HSG}) by a Legendre transform \cite{Giamarchi}  ($\beta = 1/k_B T$):
\begin{eqnarray}
\label{Eq:Sthetam}
\mathcal{S} &=& \mathcal{S}[\theta^-] + \mathcal{S}[A] + ...,  \\
\mathcal{S}[\theta^-] &=& \frac{K^-}{2\pi} \int dx \int_0^\beta d\tau \left[ \frac{1}{v^-} (\partial_\tau \theta^-)^2 + v^- (\nabla \theta^-  + A_\parallel^-)^2 \right]
\nonumber \\ 
&&\;\;\;\;- 2 g \sqrt{n_0^1 n_0^2} \int dx \int_0^\beta d\tau \cos( -\sqrt{2} \theta^- + A_\perp),\nonumber \\
\mathcal{S}[A] &=& \int dx \int_0^\beta d\tau \left[ (\text{curl} A)^2 + (\partial_\tau A)^2 \right]. \nonumber
\end{eqnarray}  
We will not require the $\phi^+$-dependent part of the action, hence the ellipsis in Eq.~(\ref{Eq:Sthetam}). For the Maxwell part of the action, $\mathcal{S}[A]$, we assume that appropriate dimensionful constants are absorbed in the derivatives.

Let us assume that quantum fluctuations are suppressed, amounting to neglecting contributions in $\partial_\tau \theta^-$ or $\partial_\tau A$. This assumption is founded if the temperature is large. The saddle point of the action corresponds to the classical ground state. The saddle point condition $\delta S / \delta \theta^-  = 0$ implies
\begin{eqnarray}
\nabla \theta_\textit{sp}^- &=& - A_{\parallel}^- \nonumber \\
\theta_\textit{sp}^- &=& \frac{1}{\sqrt{2}} a' A_{\perp}.
\end{eqnarray}
At the saddle point $A$ is constrained to be a (lattice) gradient of the arbitrary scalar function $\theta_{\textit{sp}}$.

Next, replace everywhere in Eq.~(\ref{Eq:Sthetam}) the fluctuating field $\theta^-$ by its saddle point value $\theta^-_\textit{sp}$. This is justified if $g$ and the bandwidth $v^-K^- \sim a t$ are large. To obtain the resulting action for the external gauge field, perform the gauge transformation Eqs~(\ref{Eq:AparGT}),~(\ref{Eq:AperpGT}) with scalar $f^\alpha(x) = \theta^\alpha_{\textit{sp}}$ \cite{vanWezel2007}. The saddle point action becomes
\begin{eqnarray}
\mathcal{S} &=& \frac{v^- K^- }{2\pi} \beta \int dx (\tilde{A}_\parallel^- )^2 - 2 g \sqrt{n_0^1 n_0^2} \beta \int dx  \cos(a' \tilde{A}_\perp ) \nonumber \\
&&\;\;\;\;+ \beta \int dx  (\text{curl} \tilde{A})^2 .
\end{eqnarray}
The Maxwell term does not change under the gauge transformation, however the action now contains mass terms which lead to a gapped dispersion of the modes of $\tilde{A}$. This result would have been analogously obtained by integrating out the gapped $\theta^-$ field, but the approach above (see \cite{vanWezel2007}) is less tedious.

Our treatment of the ladder Meissner effect in a continuum limit is reminiscent of the Meissner state due to phase coherence across a long Josephson junction \cite{LebwohlStephen, *FetterStephen}.  

\subsection{\MI}
\label{Subsec:MI}
We now address the emergence of Mott behavior in the \Me state. We are interested in odd mean particle number per rung, i.e.
\begin{equation}
\label{Eq:TotalDensity}
n_0^1 + n_0^2 = \frac{2 N + 1}{a},\; N \in \mathbb{N}.
\end{equation}
The simplest value is $N=0$, leading to a half-filled boson ladder, with one particle every two sites. If Eq.~(\ref{Eq:TotalDensity}) holds,  Eq.~(\ref{Eq:HSG}) becomes
\begin{eqnarray}
\label{Eq:HSGMM}
\mathcal{H}_{SG} = - 2 g \sqrt{n_0^1 n_0^2} \int dx && \; \cos( - \sqrt{2} \theta^-  + \chi x ) \times  \\
&&\left[ 1 + 2\cos\left( \sqrt{8} \phi^+ \right)  \right].  \nonumber
\end{eqnarray}
We summarize the results of this section: in the absence of long ranged repulsive interactions, there is a Mott ground state only in the Tonks gas limit $U\to\infty$. It is protected by a gap which is exponentially small with respect to the Josephson coupling $g$. Away from the Tonks limit, the Mott phase is stable if finite repulsive interactions are turned on.

We now proceed to a proof of these results. Under the energy scale $\Delta^-$ we may replace $\theta^-$ by its expectation value in Eq.~(\ref{Eq:HSGMM}). Then the effective Hamiltonian at low energies $T<\Delta^-$ is
\begin{equation}
\mathcal{H}_{SG} = -4 g \sqrt{n_0^1 n_0^2} \int dx \cos( \sqrt{8} \phi^+).
\end{equation}
This term controls the Mott transition in the total density sector. Its scaling dimension is $2 K^+$. If $K^+ < 1$, then this term as well flows to strong coupling, leading to the formation of the Mott gap
\begin{equation}
\label{Eq:Deltap}
\Delta^+ \sim \Delta^- (g^+)^{1/(2 - 2 K^+)}.
\end{equation}
We defined the dimensionless quantity $g^+ = g a / v$.

The phase appearing at $T < \Delta^+$ is the Mott insulator with Meissner currents \cite{1}, \MIMe. Importantly, note that expression~(\ref{Eq:Deltap}) holds if $K^+ < 1$, which generally corresponds to repulsive interactions of long range. These can come from intrachain repulsions or from some value of $V_\perp > 0$. If $V_\perp = 0$ in Eq.~(\ref{Eq:Hbt}), then $K^+ = K$ and the Luttinger parameter $K<1$ corresponds to long range repulsion of one-dimensional bosons \cite{Giamarchi}.

At the special value $K=1$ bosons experience hard-core interactions (the infinite interaction limit of the Tonks-Girardeau gas). The sine-Gordon term $\cos(\sqrt{8}\phi^+)$ is marginal, within our approximation of renormalization group equations. Then the Mott gap turns on exponentially but is nonvanishing even if $g$ is infinitesimal \cite{Crepin2011}
\begin{equation}\label{Eq:DeltapExp}
\Delta^+ \sim \Delta^- e^{- \alpha t/g }.
\end{equation}
The Tonks-Girardeau gas has been proved experimentally \cite{Paredes2004,Kinoshita2004}.

In general $\Delta^+ \ll \Delta^-$, which requires very small measurement temperature for the observation of the Mott insulator. We also conclude that \MI exists in the Tonks limit $U\to \infty$ or if longer ranged repulsive interactions are turned on. Let us also note that for large $V_\perp,U \gg t,g$ model~(\ref{Eq:Hbt}) maps to a gauged spin-1/2 Hamiltonian describing the Mott insulator at unit filling and in this case formally $\Delta^+ \gg \Delta^-$ \cite{1}.

\subsection{\MIMe stability}
\label{Subsec:StabilityMIMe}
Assume that the conditions are met such that \MIMe is protected by a gap $\Delta^+ < \Delta^-$. We can define critical values for flux and chemical potential beyond which the Mott insulator with Meissner currents is not stable. We perform the canonical transformation $\theta^- \to \theta^- + \frac{\chi}{\sqrt{2}} x$. The resulting form of Eq.~(\ref{Eq:JosephsonSG}) will have no oscillatory phase in the sine-Gordon terms. On the other hand, Eqs.~(\ref{Eq:LLm},\ref{Eq:LLp}) will contain terms of the form $-\int dx \mu^- \nabla \theta^- - \int dx \mu^+ \nabla \phi^+$. For the gapped phase to be stable, we require that $\mu^\pm$ do not exceed the gaps $\Delta^\pm$. This results in the following critical values for field and doping 
\begin{eqnarray}
\label{Eq:StabMIMe}
\chi_c &=& \frac{\pi \sqrt{2} \Delta^-}{v^- K^-}, \nonumber \\ 
\mu^+_c  &=& \Delta^+.
\end{eqnarray} 
The \MI state is stable for $\mu^+ < \mu^+_c$. The \Me state is stable for $\chi < \chi_c$. Two transitions out of this phase are possible: 

1. for $\chi > \chi_c$, the sine Gordon term in $\theta^-$ is irrelevant and the system enters a \VL phase \cite{OrignacGiamarchi}. The transition out of the \Me phase by increasing $\chi$ is of the commensurate-incommensurate type \cite{PT1980, Schulz1980,GiamarchiSchulz1988}.  

2. If  $\mu^+ > \mu^+_c$, it is energetically favorable to add particles to the \MI state. Due to the incommensuration, this is the \SF phase. The \MI to \SF transition by variation of $\mu$ is also a commensurate-incommensurate transition. 

We conclude that \MIMe is stable to small flux and density variations, which leads to the finite domain depicted in Fig.~\ref{Fig:PDWB}.

\subsection{\Laughlin state at $\nu=\frac{1}{2}$}
\label{Subsec:La2}
In order to energetically favor the \Laughlin state, density will be allowed to deviate from odd integer filling per rung. However, this deviation will be necessarily (very close to) commensurate with the flux. Let us focus on the following terms of Eq.~(\ref{Eq:HSG}):
\begin{widetext}
\begin{eqnarray}
\nonumber
\mathcal{H}_{\textit{SG}} &=& - 2 g \sqrt{n_0^1 n_0^2} \int dx \big[  \cos( - \sqrt{2} \theta^- + \chi x )  +
 4 \cos( -\sqrt{2} \theta^- + \chi x  ) \cos( 2\pi n_0^1 x - 2 \phi^1 ) \cos( 2 \pi n_0^2 x - 2 \phi^2 )  \Big]  + ... \\
&=&  - 2 g \sqrt{n_0^1 n_0^2 } \int dx \Big\{ \cos\left[ - \sqrt{2} \theta^- + \sqrt{8} \phi^+ + \chi x - 2 \pi (n_0^1 + n_0^2) x \right]  \nonumber \\
&& \;\;\;\;\;\;\;\;\;\;\;\;\;\;\;\;\;\;\;\;\;\;\;\;\;\;\;\;\;\;\;\;\;\;\;\;\;\;\;\;\;\;\;\;\;\;\;\;\;\;\;\;\;\;\;\;\;\;\;\;\;\;\;\;\;\;
+ \cos\left[ - \sqrt{2} \theta^- - \sqrt{8} \phi^+ + \chi x + 2 \pi (n_0^1 + n_0^2) x \right]    \Big\} +... \label{Eq:HSGforL}
\end{eqnarray}
The ellipsis in the first row represents terms containing only one of the factors $\cos( 2\phi^1 - 2\pi n_0^1 x)$, $\cos( 2\phi^2 - 2\pi n_0^2 x)$, $\cos[ \sqrt{8}\phi^- - 2\pi (n_0^1- n_0^2) x]$. We assume that $n_0^{1} \approx n_0^2 \approx (2N+1)/(2a)$, where the approximate equality is such that all three factors are oscillatory and can be discarded. In the second row, we assume that $\chi > \chi_c$ such that $\cos(-\sqrt{2}\theta^- + \chi x)$ can be discarded as explained in Sec.~\ref{Subsec:StabilityMIMe}.

Our purpose now is to tune flux and density such that one of the two terms in Eq.~(\ref{Eq:HSGforL}) stays relevant. The oscillatory argument in the first or second term of Eq.~(\ref{Eq:HSGforL}) vanishes if the following commensuration condition holds 
\begin{equation}
\label{Eq:LaCs}
a\left[ 2 \pi (n_0^1 + n_0^2) \pm \chi \right] = 0\text{ mod } 2\pi. 
\end{equation}
Note that if the system is gapless any change in chemical potential results in doping $\frac{v^+}{\pi K^+} \delta^+ = -\mu^+$. Therefore condition~(\ref{Eq:LaCs}) can be attained within the gapless \SFVL by continuously changing flux and chemical potential.

If we pick the lower sign for Eq.~(\ref{Eq:LaCs}), Eq.~(\ref{Eq:HSGforL}) reduces to
\begin{eqnarray}
\label{Eq:LaughlinCouplingm}
\mathcal{H}_{\textit{SG}} = - 2 g \sqrt{n_0^1 n_0^2}  \int dx \cos( -\sqrt{2} \theta^- + m \sqrt{2} \phi^+ ),\;\; m = 2.
\end{eqnarray}
For a general integer $m$, Eq.~(\ref{Eq:LaughlinCouplingm}) represents the correlated hopping term in the coupled chain construction of the $\nu =\frac{1}{m}$ Laughlin state \cite{Kane2002,TeoKane}. Terms with $m > 2$ have larger scaling dimension. In the following, we will provide some results as a function of $m$ for generality.  
\end{widetext}

In the case $m = 2$, the scaling dimension of Eq.~(\ref{Eq:LaughlinCouplingm}) is $\delta = 1/(2K^-) + 2 K^+$. At $V_\perp = 0$ this term is irrelevant unless long ranged interactions along the chains are present. The coupling constant $g$ can become relevant in the presence of sufficiently large $V_\perp > 0$. The associated energy gap is
\begin{equation}
\label{Eq:DelLa}
\Delta \sim g \left( \frac{g a}{v} \right)^{\frac{1}{2 - 1/(2K^-) - 2 K^+}}.
\end{equation}
When the coupling is marginal, the gap has an exponential dependence on $g$ as in Eq.~(\ref{Eq:DeltapExp}). To summarize, the \Laughlin state should be observable with sufficiently strong repulsive interactions $V_\perp$ (or sufficiently long range repulsive interactions along the chains). Otherwise, when the coupling constant $g$ is irrelevant, a gap will still be observable in finite sized systems for a sufficiently strong bare value.

The addition of a particle spoils the commensuration between mean density and flux, and therefore the term of Eq.~(\ref{Eq:LaughlinCouplingm}) becomes gapless. However, a finite chemical potential is required to add an extra particle to the system. We arrive therefore at stability conditions analogous to Eqs.~(\ref{Eq:StabMIMe}). A ``surplus'' chemical potential $\delta \mu^+$ causing deviations in mean density from the background density which satisfies Eq.~(\ref{Eq:LaCs}) must not exceed $\Delta$
\begin{equation}
\delta \mu^+ < \delta \mu^+_c = \Delta.
\end{equation}
Moreover, a ``surplus'' flux that causes deviations from Eq.~(\ref{Eq:LaCs}) must obey
\begin{equation}
\delta \chi < \delta \chi_c = \frac{\pi \sqrt{2} \Delta}{v^- K^-}.
\end{equation} 
A sketch of the possible region following from the present discussion is depicted on Fig.~\ref{Fig:PDWB}B. However, the clear delimitation of such a region depends on the details of the microscopic Hamiltonian. Note that another possibility would be to pick the upper sign in~(\ref{Eq:LaCs}), yielding a state related to the one discussed above by particle-hole symmetry.


\section{Observables}
\label{Sec:Obs}
In this section we discuss various observable quantities that allow us to characterize the phases \MIMe and \Laughlin. We begin with a definition of the lattice current operators and a lattice version of the flux quantization condition obtained from a Gross-Pitaevskii approximation of the boson operator, in Sec.~\ref{Subsubsec:Current}. We continue to a discussion of current operators (\ref{Subsubsec:CurrentOpMJ}), flux quantization (\ref{Subsubsec:FluxQuMJ}) and Hall responses from a Laughlin argument (\ref{Subsubsec:LaughlinArgMJ}) in \MIMe. The analogous discussion for \Laughlin appears in \ref{Subsubsec:CurrentOpL}. We discuss the gapless effective edge model of \Laughlin in \ref{Subsubsec:EdgeTheory}.

\subsection{Current operator and lattice flux quantization}
\label{Subsubsec:Current}
\begin{figure}
\includegraphics[width=\linewidth]{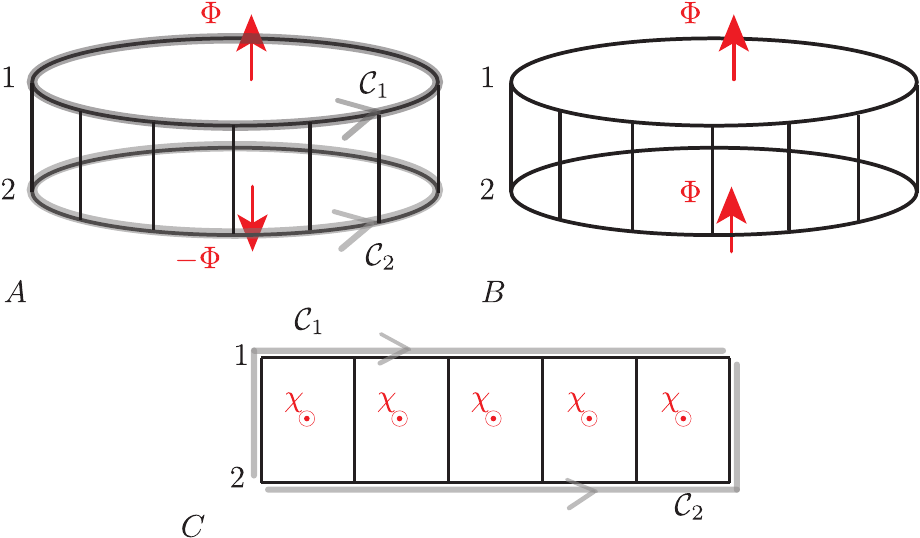}
\caption{\label{Fig:FluxQu}Two leg ladder. \textbf{A:} Periodic boundary conditions. Closed loops $\mathcal{C}_1$ and $\mathcal{C}_2$ traverse chains 1 and 2 in the counter-clockwise direction. $\Phi = L \chi/2$ is the net flux per chain. \textbf{B:} An Aharonov-Bohm flux through the periodic geometry. \textbf{C:} The same flux can be achieved with open boundary conditions by threading flux $\chi$ per plaquette. }
\end{figure}
The flux quantization condition for a superfluid \cite{Thouless} relates the winding number of the boson phase around closed loops on the lattice with current circulation and flux. We begin by fixing the definition for lattice current. Assuming a generic quadratic Hamiltonian in the form $H = \sum_{ij} |t_{ij}|e^{i \phi_{ij}} b^\dagger_i b_j$, the current operator is obtained from the Heisenberg equation of motion $\dot{n}_i = i[H, n_i] \equiv \sum_{j} j_{ij}$, where
\begin{equation}
j_{ij} = - i |t_{ij}| e^{i\phi_{ij}} b^\dagger_i b_j + \text{H.c.}  
\end{equation}
For a bond $b \equiv ij$, where $i$ and $j$ denote sites, the current operator $j_{ij}$ measures the number of particles per unit time flowing from site $j$ into site $i$. Consider now a loop on the lattice, denoted by the sequence of bonds $\mathcal{C} = b_0, b_1, ...$ . In a superfluid, writing boson creation operators as $b_i^\dagger \approx \sqrt{n_0} e^{- i\theta_i}$, neglecting density fluctuations, and expanding for small gauge invariant phases $-\theta_i + \theta_j + \phi_{ij}$, we obtain the following condition for any closed loop $\mathcal{C}$ on the lattice.
\begin{equation}
\label{Eq:QuantCond}
\sum_{b \in \mathcal{C}} \frac{1}{2 |t_b| n_0} \langle j_b \rangle + \Phi_\mathcal{C} = 2 \pi N_\mathcal{C}.
\end{equation}
We have defined $t_b$ as the hopping integral on bond $b$. The phase $\Phi_\mathcal{C} = \sum_{b\in \mathcal{C}} \phi_{b}$ corresponds to the line integral of the gauge field along curve $\mathcal{C}$. The winding number of the boson phase field around $\mathcal{C}$ is denoted by the integer $N_\mathcal{C}$. Eq.~(\ref{Eq:QuantCond}) is the flux quantization condition for a lattice, which is consistent with the continuum result \cite{Thouless}.

\subsection{Current operators in \MIMe}
\label{Subsubsec:CurrentOpMJ}
We begin by computing the current expectation values in \MIMe. The boson current operator is obtained from the following Heisenberg equation
\begin{equation}
i [ \mathcal{H}, -\frac{1}{\pi} \sqrt{2} \nabla \phi^-(x) ] = \frac{d}{dt} (n_1 - n_2),
\end{equation}
which can be separated into interchain and intrachain current operators
\begin{eqnarray}
\label{Eq:BMMC}
j_\perp &=& 2 \frac{g}{\pi a}  \sin \left(\sqrt{ 2 }\theta^- - \chi x \right), \nonumber \\ 
j_\parallel &=&  j_\parallel^1 - j_\parallel^2 = - v^- K^-  \sqrt{2}\nabla \theta^- .
\end{eqnarray}
Whenever the relative phase between the chain condensates is pinned, \textit{i.e.} for energy scales below $\Delta^-$, we find
\begin{eqnarray}
\label{Eq:bmc}
\langle j_\perp(x) \rangle &=& 0, \\ \nonumber
\langle j_\perp(x) j_\perp(0) \rangle_\textit{connected}  &=& 0 \nonumber \\
\; \langle j_\parallel(x) \rangle &=& - a t \chi, \label{Eq:bmc2}
\end{eqnarray}
that is, ``bulk'' currents and their fluctuations vanish, whereas equal counterflowing ``edge'' currents have a difference $ - 2 t a^2 n_0 \chi $, where $n_0 = \frac{1}{2a}$ is the mean boson density per chain. We remark that the form of the current operator expectation value, Eq.~(\ref{Eq:bmc2}), changes in the strong coupling limit $U,V_\perp \gg t,g$, where \cite{1} at second order in perturbation theory $j_\parallel(x) = - 2 a (t^2/V_\perp) \chi$. In either the weak or the strong coupling regime, current $j_\parallel$ persists in the Mott phase, where the field $\phi^+$ is pinned to the classical value $\langle \phi^+ \rangle = 0$. In addition, this phase exhibits vanishing density fluctuations 
\begin{equation}
\langle (n^1 + n^2)(x) (n^1 + n^2)(0)\rangle_\textit{connected} = 0,
\end{equation}
and a density pattern with one boson per rung.

\subsection{Flux quantization in \MIMe}
\label{Subsubsec:FluxQuMJ}
The current expectation values obtained in Eq.~(\ref{Eq:BMMC}) obey a flux quantization condition similar to Eq.~(\ref{Eq:QuantCond}). Consider that the loop is the boundary of the cylinder formed by the two periodic chains. In Figure~\ref{Fig:FluxQu}A this is $\mathcal{C} = \mathcal{C}_1 - \mathcal{C}_2$, where the minus sign means that $\mathcal{C}_2$ is traversed in reverse. Then, using Eq.~(\ref{Eq:BMMC}) in (\ref{Eq:QuantCond}) we obtain the following expression
\begin{eqnarray}
\label{Eq:CylQuantCond}
2\pi N_{\mathcal{C}_1 -\mathcal{C}_2} &=& \frac{1}{at} \int_0^{La} dx \langle j_\parallel^1 - j_\parallel^2 \rangle   + L a \chi  = 0.  
\end{eqnarray}
The second equality follows from~(\ref{Eq:bmc2}). The vanishing winding number of the superfluid phase is a signature of the Meissner effect. 

Let us now realize flux $\chi$ per plaquette via the choice $A^1_{j,j+1} = - A_{j,j+1}^2 = \chi/2$ in Eq.~(\ref{Eq:Hbt}). With this choice, there is no net flux parallel to the axis of the cylinder, as depicted in Figure~\ref{Fig:FluxQu}A. Writing the condition in Eq.~(\ref{Eq:QuantCond}) for each path $\mathcal{C}_1$ and $\mathcal{C}_2$ and using Eq.~(\ref{Eq:CylQuantCond}), we find 
\begin{eqnarray}
\label{Eq:ChainQuantCond}
2\pi N_{\mathcal{C}_1} &=& \frac{1}{at} \int_0^{La} dx \langle j_\parallel^1 \rangle + L a \chi/2 \nonumber \\ &=& 2\pi N_{\mathcal{C}_2} = \frac{1}{at} \int_0^{La} \langle j_\parallel^2 \rangle - L a \chi/2.
\end{eqnarray}
Eq.~(\ref{Eq:ChainQuantCond}) represents the flux quantization condition in our setup.  While Eq.~(\ref{Eq:ChainQuantCond}) holds for periodic boundary conditions, little changes qualitatively for open boundaries, with the loop $\mathcal{C}_1 - \mathcal{C}_2$ as in Figure~\ref{Fig:FluxQu}C. For a loop surrounding the ladder, there will be $O(1/L)$ corrections, due to rung currents appearing at the open boundaries. In general, the flux quantization provides a way to measure the winding of the phase of the boson wavefunction from current measurements, and detect the presence of vortices in the sample.

\subsection{$\sigma_{xy}$ from Laughlin argument in \MIMe}
\label{Subsubsec:LaughlinArgMJ}
The Hall response is vanishing in \MIMe. Suppose a current $j^1_{\parallel} = j^2_\parallel$ is generated along the horizontal direction in Figure~\ref{Fig:Ladder}. A ``voltage drop'' to realize such a current can be realized by tilting both chains in the same direction, or by adiabatically threading a flux quantum between the chain ends as depicted in Figure~\ref{Fig:FluxQu}B \cite{Laughlin1981,*Halperin1982}. If $\sigma_{xy} \neq 0$, the response to this must be a perpendicular flow of current amounting to a quantized charge after a full period \cite{Thouless1983}. There is a converse situation, a tilt between the chains would cause a uniform $\langle j^{1}_\parallel + j^{2}_\parallel \rangle \neq 0$. However, since from Eq.~(\ref{Eq:BMMC}) the commutator of operators $j_\perp$ and $j_{\parallel}^{1,2}$ vanishes, the Kubo formula for the conductivity implies
\begin{equation}
\sigma_{xy} = 0.
\end{equation}
This result is expected as the phase \MIMe is fully gapped.

\subsection{Current operators in \Laughlin}
\label{Subsubsec:CurrentOpL}
Similarly, we compute bosonic particle currents in the phase \Laughlin. The plaquette currents of the bosonic particles are 
\begin{eqnarray}
\label{Eq:LaC}
j_\perp &=& 2 \frac{g n_0}{\pi}  \sin \left(\sqrt{ 2 }\theta^- - m\sqrt{2} \phi^+ \right), \nonumber \\ 
\;j_\parallel &=&  j_\parallel^1 - j_\parallel^2 = - v^- K^-  \sqrt{2}\nabla \theta^-.
\end{eqnarray}
These lead to the following expectation value in the gapped phase
\begin{eqnarray}
\langle j_\perp \rangle = 0.
\end{eqnarray}
Bulk currents vanish. Assuming that total charge fluctuations vanish, the circulation of current along the contour $\mathcal{C}_1 - \mathcal{C}_2$ in Figure~\ref{Fig:FluxQu} vanishes:
\begin{equation}
\label{Eq:CircLa2}
\frac{1}{t}\int_0^{La} dx \langle j_\parallel \rangle = \frac{1}{t}\left[\int_0^{La} dx \langle j^1_\parallel \rangle - \int_0^{La} dx \langle j^2_\parallel \rangle \right] = 0.
\end{equation} 
The contrast between Eq.~(\ref{Eq:CircLa2}) and Eq.~(\ref{Eq:CylQuantCond}) can be used to distinguish \Laughlin from \MIMe. In addition, this phase is not fully gapped, as described below.

\subsection{Chiral edge modes in \Laughlin}
\label{Subsubsec:EdgeTheory}
In this subsection we discuss the structure of the edge theory when the bulk term of Eq.~(\ref{Eq:LaughlinCouplingm}) produces a gap. Let us define \cite{Kane2002,TeoKane} new chiral fields in the form 
\begin{equation}
\label{Eq:ChiralFields}
\phi_{r}^\alpha = \theta^\alpha/m + r \phi^\alpha
\end{equation}
for left ($r=-1$) and right ($r=+1$) moving excitations in chain $\alpha = 1,2$. The following commutation relations follow from the algebra in Eq.~(\ref{Eq:A}), 
\begin{equation}
\label{Eq:ChiKMAlg}
[ \phi_r^\alpha(x), \phi_{r'}^\beta(x') ] = i r \frac{\pi}{m} \delta_{rr'} \delta_{\alpha\beta} \text{Sign}(x'-x).
\end{equation}
The algebra of the chiral modes in Eq.~(\ref{Eq:ChiKMAlg}) implies that the momentum associated with $\phi_r^\alpha (x)$ is 
\begin{equation}
\label{Eq:ChiMom}
\Pi_r^\alpha(x) \equiv \frac{m}{2\pi r} \nabla \phi_r^\alpha(x).
\end{equation}
Let us define new density and phase fields for the bulk, respectively:  
\begin{eqnarray}
\label{Eq:BulkFields}
\phi = (-\phi^1_{-1} + \phi^2_{+1})/2, \theta = (\phi^1_{-1} + \phi^2_{+1})/2.
\end{eqnarray}
These are related to the original fields $\theta^\pm, \phi^\pm$ through
\begin{eqnarray}
\phi = -\frac{\sqrt{2}}{2m}\theta^- + \frac{\sqrt{2}}{2} \phi^+,\; \theta = \frac{\sqrt{2}}{2m}\theta^+ - \frac{\sqrt{2}}{2} \phi^-.
\end{eqnarray}
These fields obey the Kac-Moody algebra
\begin{equation}
[\phi(x) , \theta(x')] = i \frac{\pi}{2m} \text{Sign}(x' - x).
\end{equation}
Using the new bulk variables we define the ``bulk'' charge density $n = -\frac{1}{\pi} \nabla \phi$, whereas the quasiparticle density is given by $n_{\text{QP}} = -\frac{m}{\pi} \nabla \phi$. A kink of $2\pi$ in the field $2m\phi$ corresponds to the creation of one Laughlin quasiparticle. The correlated hopping term Eq.~(\ref{Eq:LaughlinCouplingm}) pins the left chiral field of chain 1 to the right chiral field of chain 2:
\begin{equation}
\label{Eq:LaughlinCouplingPhi}
\mathcal{H}_\textit{SG} = - 2 g n_0 \int dx \cos( 2m \phi ).
\end{equation}
The full Hamiltonian is given by 
\begin{equation}
\label{Eq:FullHLaughlinCoupling}
\mathcal{H}[ \phi(x), \theta(x), \phi^{1}_{+1}(x), \phi^{2}_{-1}(x) ] = \mathcal{H}_0^+ + \mathcal{H}_0^- + \mathcal{H}_\textit{SG}.
\end{equation}
with $\mathcal{H}_\textit{SG}$ specified in Eq.~(\ref{Eq:LaughlinCouplingPhi}) and $\mathcal{H}_0^\pm$ as in Eqs.~(\ref{Eq:LLm},\ref{Eq:LLp}). 

It is possible to obtain the effective low energy theory of the remaining two gapless chiral modes $\phi_{+1} \equiv \phi^1_{+1}$ and $\phi_{-1} \equiv \phi^2_{-1}$. The detailed calculation is given in Appendix~\ref{Ap:EdgeTheory}. In summary, the result of integrating out the massive field $\phi$ is a generic Luttinger liquid 
\begin{eqnarray}
\label{Eq:Ledge}
4 \pi \mathcal{L}_\textit{edge} = \int dx \left(  \dot{\phi}_r K_{rr'} \nabla \phi_{r'} - \nabla \phi_r V_{rr'} \nabla \phi_{r'}    \right).
\end{eqnarray} 
The matrix $K_{rr'} = m \, r \delta_{rr'}$ is determined by Eq.~(\ref{Eq:ChiMom}). It describes two counterpropagating modes on distinct edges of the Laughlin state $\nu = \frac{1}{m}$ \cite{WenZee1992, Wen1995}.

Note however that Eq.~(\ref{Eq:Ledge}) does \textit{not} describe a chiral Luttinger liquid. The matrix $V_{rr'}$ is nonuniversal and has nonvanishing off-diagonal elements. These terms describe backscattering between the chiral fields. We provide the explicit form of $V_{rr'}$ in Appendix~\ref{Ap:EdgeTheory}. The resulting Luttinger parameter $K_\textit{edge}$ engenders a more complex charge fractionalization phenomenon \cite{SafiSchulz,*Pham2000,*KHY,*BergEtAl}. Interestingly, when backscattering terms between the edges are suppressed, current noise between the edges probes the fractional charge of bulk quasiparticles \cite{KaneFisherNoise}. This has been demonstrated experimentally \cite{dePicciotto,*Radu}, in the two-dimensional electron gas. Such experiments are feasible as well in ultracold atom systems, where quantum point contacts have already been realized \cite{Brantut2012,*Brantut2013,*Krinner2014}.

Let us consider briefly the case of $N$ chains with the same coupling Eq.~(\ref{Eq:LaughlinCouplingm}) between consecutive chains. Backscattering terms between the chiral edge modes, $\phi_{+1}^1$ and $\phi_{-1}^N$ vanish exponentially fast with $N$. The remaining action describes the edge degrees of freedom. It consists of a chiral Luttinger liquid 
\begin{equation}
4 \pi \mathcal{L}_\textit{edge} = \sum_{r=\pm 1} \int dx\left[ \; m r \dot{\phi}_r  \nabla \phi_{r} - v (\nabla \phi_r)^2 \right].
\end{equation}
If the bulk were a continuous two dimensional manifold, this edge theory would correspond to the bulk Chern Simons theory \cite{Wen1992}
\begin{equation}
4 \pi S[ \mathcal{A} ] = \frac{1}{m} \int dx dy dt \; \epsilon^{\mu\nu\rho} \delta\mathcal{A}_\mu \partial_\nu \delta\mathcal{A}_\rho.
\end{equation}
for the external gauge field. $\delta\mathcal{A} = \mathcal{A} - A$ is the deviation of the dynamical gauge field from the fixed background field $A$. $\epsilon^{\mu\nu\rho}$ is the Levi-Civita symbol. The relation between this bulk Chern-Simons theory and the edge theory is established by requiring that the action defined over a two-dimensional manifold with boundary be gauge invariant.  

We stress again that the underlying assumption of this discussion was that~(\ref{Eq:LaughlinCouplingm}) is relevant. The coupling $g$ is relevant for sufficiently strong $V_\perp$, across rungs between consecutive chains, or sufficiently long ranged intrachain repulsive interaction. However, in finite sized systems, a gap can still be associated with~(\ref{Eq:LaughlinCouplingm}) if $g$ is larger or comparable to the energy scale of the bandwidth, set by the intrachain hopping $t$.


\section{Hybrid fermion-Cooper pair analogues}
\label{Sec:WF}
\begin{figure}[t!]
\includegraphics[width=0.90 \linewidth]{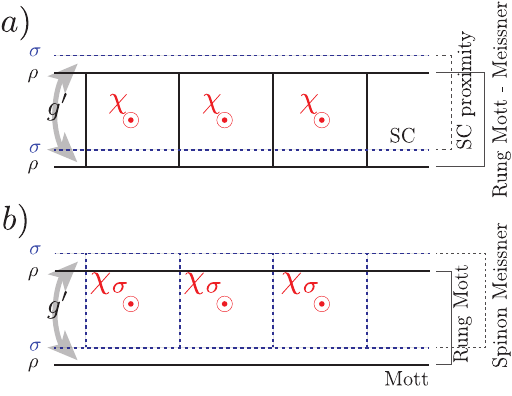}
\caption{\label{Fig:FermiLadder}Possible coupling schemes for a fermion ladder: a) Cooper pair \MIMe is realized by starting with superconductivity in 1. The proximity effect induces superconductivity in chain 2, via a term \SC (see Table~\ref{Tab:PFL}). At lower energy scales, the \MIMe (or depending on filling and flux any other state of Table~\ref{Tab:PFL}) forms. b) Dual model: Starting with MI in chain 1, the proximity effect makes chain 2 insulating, via \MI. At lower energy scales, the spin sector is in the phase \SMe.}
\end{figure}

\begin{table}
\begin{tabular}{|l|l|l|}
\hline
Sector & Notation & Phase description                \\
       &          & sine-Gordon term                 \\
\hline
$\rho$,+      & \MI   &  $2\phi_\rho^+$  \\
$\rho$,+      & \SF   &  $2\phi_\rho^+ + 2(k_F^1 + k_F^2) x$    \\
$\rho$,-      & \Me   &  $2\theta_\rho^-$          \\
$\rho$,-      & \VL   &  $2\theta_\rho^- - 2\chi x$     \\
$\rho$,+\& -  & \Laughlin      & $2\theta_\rho^- - 2\phi_\rho^+$        \\
\hline
$\sigma$,+    & \SC   &  $2\phi_\sigma^+$  \\
$\sigma$,-    & \SMe  &  $2\theta_\sigma^-$  \\
$\sigma$,+\&- & \Laughlin      &  $2\theta_\sigma^- - 2\phi_\sigma^+$  \\
\hline   
\end{tabular}
\caption{\label{Tab:PFL} Phases of the Josephson ladder 
appearing in the phase diagram of Fig.~\ref{Fig:PDWB}. The
``+/-'' sector denotes total/relative vertical bond (rung) density $n_i^1 +/- n_i^2$ 
(see Fig.~\ref{Fig:Ladder}). The ``+'' sector can be in a Mott insulator
or superfluid phase, whereas the ``-'' sector can be in a Meissner phase or 
a vortex lattice phase depending on the strength of the field. The Laughlin
phase arises from a condition that mixes the two sectors. } 
\end{table}

In this section we prove that spinful fermions on the lattice of Fig.~\ref{Fig:Ladder} have ground states analogous to those of bosons presented in Sec.~\ref{Sec:WB}. This model, in the absence of gauge fields has been introduced in Ref. \cite{Karyn2001} to (qualitatively) describe the pseudo-gap phase of high-Tc superconductors (modeled as ``hot spots'' with preformed Cooper pairs and ``cold spots'' Fermi arcs \cite{Rice2012, *Schulz1987, *ZanchiSchulz1998, *HalbothMetzner2001, *Honerkamp2001, BalentsFisherNayak, Geshkenbein1997, *IoffeMillis1998, *Geshkenbein1998}). We note that recently novel features in high-Tc superconductors with magnetic fields or in relation with density wave order have been discussed \cite{Marcel,*Seamus,*Sachdev,*Pepin,*TsvelikChubukov,TailleferProust1,*TailleferProust2,*TailleferProust3}. 

There exists a rich literature on the topic of two-leg fermion ladders \cite{CarrNarozhnyNersesyanAnnPhys2013}. For spinless fermions, a striking phenomenon in the presence of magnetic field is the existence of the orbital antiferromagnetic phase, (also called d-density wave  \cite{Kotliar1988}, or staggered flux phase of the mean-field Hamiltonian proposed by Affleck and Marston \cite{AffleckMarston1988, *MarstonAffleck1989} introduced in the context of high-$T_c$ superconductivity) coexisiting with the bond-density wave \cite{NarozhnyCarrNersesyan,*CarrNarozhnyNersesyan,*Schollwoeck2003}. Carr and Tsvelik \cite{CarrTsvelik} studied the model of spin-gapped chains coupled by Josephson terms in magnetic fields and found competing charge density wave and superconducting correlations. Roux \textit{et al.} \cite{Roux2007} have studied the magnetic orbital effect in doped two-leg spinful fermionic ladders and found a reentrant transition into a spin gapped phase at high magnetic flux.

We summarize the main results in this section: If superconducting correlations dominate, ground states analogous to those of Sec.~\ref{Sec:WB} occur. The Cooper pair \MIMe ground state is separated by a finite gap from the rest of the spectrum. This gap depends on the interchain coupling $g$ like a power law even when long range repulsive interactions are off. If charge density wave correlations dominate, as happens when each chain is at half filling, then superfluidity and the Meissner effect can occur in the spinon sector. To aid throughout the discussion, Table~\ref{Tab:PFL} lists the phases encountered in this section, along with the relevant charge or spin sector, and terms in the Hamiltonian inducing the particular order.

\begin{figure}
\includegraphics[width=0.90 \linewidth]{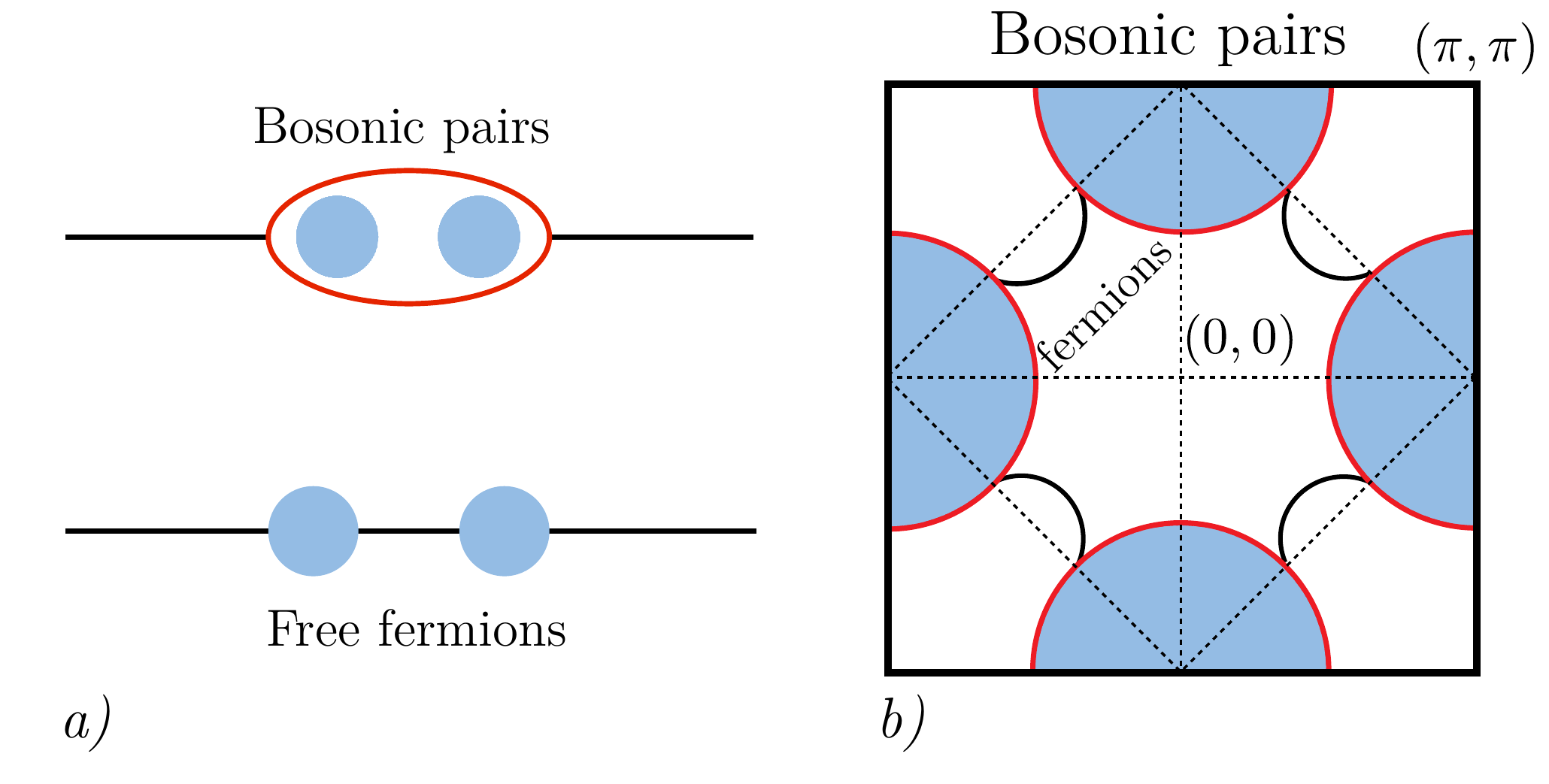}
\caption{Hybrid two-leg ladder system introduced in Ref. \onlinecite{Karyn2001} with preformed Cooper pairs in one chain and repulsive fermions in the other chain. Here, we extend the model by discussing magnetic field effects. This toy model shows certain analogies with the pseudo gap phase of high-Tc cuprates showing hot spots (preformed Cooper pairs) and cold spots (Fermi arcs) \cite{Rice2012, *Schulz1987, *ZanchiSchulz1998, *HalbothMetzner2001, *Honerkamp2001, BalentsFisherNayak, Geshkenbein1997, *IoffeMillis1998, *Geshkenbein1998}.}
\end{figure}

We start with a microscopic tight-binding model of spinful fermions with Hubbard interactions on the same ladder lattice of Fig.~\ref{Fig:Ladder}.
\begin{eqnarray}
\label{Eq:HLat}
H &=& H_1 + H_2 + H_\perp \nonumber  \\
H_\alpha &=& \sum_{\sigma i} \left[ -t c^\dagger_{\alpha \sigma}(i)  c_{\alpha \sigma}(i+1) + \text{H.c.} \right] \nonumber \\ 
&&+ \sum_{i} U_\alpha  n_{\alpha\uparrow}(i) n_{\alpha\downarrow}(i)  \nonumber \\
H_\perp &=& \sum_{\sigma i} \left[ -g e^{ i a' A_{\perp i} }\, c^\dagger_{1\sigma}(i) c_{2 \sigma}(i) + \text{H.c.} \right].
\end{eqnarray}
The operator $c^\dagger_{\alpha \sigma}(i)$ creates a fermion of spin $\sigma = \uparrow$ or $\downarrow$ on chain $\alpha=1$ or $2$ at site $i$, and $n_{\alpha \sigma}(i) = c^\dagger_{\alpha \sigma}(i) c_{\alpha \sigma}(i)$ is the fermion number operator. In Eq.~(\ref{Eq:HLat}) we consider periodic boundary conditions. 

To study the possible phases of this system it is again convenient to use bosonization. We assume that interchain coupling is small. The basis suitable for expressing the continuum Hamiltonian consists of field operators $\psi^\alpha_{\sigma}(x) = c_{\alpha \sigma}(j)/\sqrt{a}$.  In the free particle model, we assume that Fermi momenta $k_F^\alpha$, the coupling $g$, and the flux per plaquette $\chi$ are chosen such that the Fermi surface contains two points. We approximate the fermion field operator as a sum over right-moving and a left-moving contributions at the two Fermi points,
\begin{equation}
\label{Eq:FermiField}
\psi^\alpha_{\sigma}(x) = \psi^{\alpha}_{+,\sigma}(x) + \psi^{\alpha}_{-,\sigma}(x).
\end{equation}
Next, it is necessary to introduce bosonic fields describing charge degrees of freedom, $\phi^{\alpha}_{\rho}(x)$, $\theta^{\alpha}_{\rho}(x)$; and spin degrees of freedom, $\phi^{\alpha}_{\sigma}(x)$, $\theta^{\alpha}_{\sigma}(x)$. Each pair of operators obeys the algebra in Eq.~(\ref{Eq:A}), and either one of the $\rho$ fields commutes with the $\sigma$ fields.  The chiral fermion operators in Eq.~(\ref{Eq:FermiField}) are \cite{Giamarchi}:
\begin{equation}
\label{Eq:FermiFieldAlpha}
\psi^\alpha_{r,s} (x) = \frac{U^\alpha_{r,s}}{\sqrt{2\pi a}}  e^{ i r k^\alpha_{F} x} e^{- \frac{i}{\sqrt{2}} \left[  r \phi^\alpha_{\rho}(x) - \theta^\alpha_{\rho} (x) + s (r \phi^\alpha_{\sigma}(x) - \theta^\alpha_\sigma)  \right]} 
\end{equation}
for $\alpha = 1,2$, $r = \pm 1$, and $s = \pm 1$. $U_{r,s}$ are Klein factors enforcing Fermi statistics. For our purposes it is sufficient to neglect their contribution and replace them by unity. We have denoted the Fermi momentum in chain $\alpha$ by $k_F^\alpha$. The Hamiltonian density corresponding to Eq.~(\ref{Eq:HLat}) is
\begin{eqnarray}
\label{Eq:HFb}
\mathcal{H} &=& \mathcal{H}_1 + \mathcal{H}_2 + \mathcal{H}^I_{\perp} \nonumber \\
\mathcal{H}_\alpha &=& \mathcal{H}^\alpha_{0\rho} + \mathcal{H}^\alpha_{0\sigma} + \frac{U_\alpha}{2\pi^2 a} \int dx \cos( \sqrt{8} \phi^\alpha_\sigma ).
\end{eqnarray}
The Hamiltonians $\mathcal{H}_{0\rho}^\alpha$ and $\mathcal{H}_{0\sigma}^\alpha$ represent Luttinger liquids [see Eq.~(\ref{Eq:LLp})] with sound velocities and Luttinger parameters given by \cite{Giamarchi}:
\begin{eqnarray}
\label{Eq:KsUs}
K^\alpha_\rho = 1/\sqrt{ 1 + \frac{U_\alpha a}{\pi v_F^\alpha}} &,&\, K^\alpha_\sigma = 1/\sqrt{ 1 - \frac{U_\alpha a}{\pi v_F^\alpha}}, \nonumber \\
v^\alpha_\rho = v_F^\alpha  \sqrt{ 1 + \frac{U_\alpha a}{\pi v_F^\alpha}} &,&\, v^\alpha_\sigma = v_F^\alpha \sqrt{ 1 - \frac{U_\alpha a}{\pi v_F^\alpha}}.
\end{eqnarray} 

The tunneling term $\mathcal{H}^I_\perp$ in Eq.~(\ref{Eq:HFb}) is given explicitly in Appendix~\ref{Ap:And}, Eq.~(\ref{Eq:IntHperp}). The essential property is that $\mathcal{H}_\perp^I$ contains terms $\propto \exp( i \theta_\sigma^{1,2})$ and is therefore irrelevant if a spin gap is open in chain $1$ or $2$ via the sine-Gordon potentials in Eq.~(\ref{Eq:HFb}). Note that since the tunneling term also contains terms $\propto e^{i\theta_\rho^{\alpha}}$, the development of a charge gap due to umklapp terms would also make the tunneling term irrelevant. For the moment, let us assume that the fermions are at half-filling, $k_F^1 + k_F^2 = \frac{\pi}{a}$, but that this condition is fulfilled while each chain is away from half-filling, such that \textit{umklapp} terms are irrelevant.

The sine-Gordon terms $U_\alpha$ are responsible for opening a spin gap $\Delta_\sigma^\alpha$ in chain $\alpha$ whenever $U_\alpha < 0$. Let us assume that the interactions in chain 1 are attractive, such that the associated coupling $g^1_{1\perp} = U_1$ flows to strong coupling and a spin gap $\Delta_\sigma^1$ opens. This phase is the Luther-Emery liquid \cite{LutherEmery}. We moreover assume that interactions in chain 2 are repulsive, such that there is no spin gap. Then the effective theory of chain 2 is described by a fixed point Luttinger liquid Hamiltonian with parameter $(K_\sigma^2)^* = 1$. For energy scales under $\Delta_\sigma^1$, interchain hopping terms $\mathcal{H}_\perp^I$ are irrelevant, and the coupling between the chains is given by terms obtained at second order in perturbation theory. One of us has shown that superconductivity is induced in chain 2 from the proximity with chain 1 of preformed pairs  \cite{Karyn2001}. 

The Hamiltonian at second order in $g$ is (see Appendices~\ref{Ap:And} and \ref{Ap:SecondOrderPT})
\begin{eqnarray}
\label{Eq:FHSG}
\mathcal{H}^{II}_\perp &=& -\frac{g'}{a}\int dx \cos\left[ \sqrt{2} \phi^{2}_{\sigma} \right] \big\{ \cos\left[ 2( k_F^1 + k_F^2 ) x -  2 \phi^{+}_\rho \right] \nonumber \nonumber \\
&&\;\;\;\;\;\;\;\;\;\;\;\;\;\;\;\;\;\;\;\;\;\;\;\;\;\;\;\;\;\;\;\;\;\;\;\;+  2\cos\left[ 2 \chi x - 2 \theta^-_\rho \right] \big\}   \\ 
&& - \frac{g'}{a} \int dx \cos\left[ 2 \chi x - 2 \theta^-_\rho \right] \cos\left[ 2( k_F^1 + k_F^2 ) x -  2 \phi^{+}_\rho \right] \nonumber,
\end{eqnarray}
where 
\begin{eqnarray}
\frac{g'}{a} &=& \frac{2 g^2}{\Delta_\sigma^1} \frac{1}{\pi a}, \\
 2 \chi x &=& a' A_{\perp\downarrow} + a' A_{\perp\uparrow},
\end{eqnarray}
with infinitesimal $\chi$ representing the flux per plaquette per spin species. 

Note that the first two of the three terms in Eq.~(\ref{Eq:FHSG}) gap the field $\phi^2_\sigma$, triggering the formation of Cooper pairs in the second chain \cite{Karyn2001}. We mark the opening of the superconducting gap in the second chain by \SC on Figure~\ref{Fig:FermiLadder}a). The coupling is represented between the blue dashed lines since it occurs in the $\sigma$ sector. 

The first term of Eq.~(\ref{Eq:FHSG}) is a spin conserving backscattering term $(\psi_{-r,\sigma}^1)^\dagger \psi_{+r,\sigma}^2 (\psi_{-r,\sigma}^2)^\dagger \psi_{+r,\sigma}^1 + \text{H.c.}$ It favors a charge density wave. Its scaling dimension is
\begin{equation}
\delta_1 = \frac{ 1 }{ 2 } + \frac{K^1_\rho}{2} + \frac{K^2_\rho}{2}.
\end{equation}

The second term of Eq.~(\ref{Eq:FHSG}) corresponds to the tunneling of Cooper pairs $(\psi_{-r,\sigma}^1)^\dagger \psi_{+r,\sigma}^2 (\psi_{+r,-\sigma}^1)^\dagger \psi_{-r,-\sigma}^2 e^{ i a' (A_{\perp \sigma} + A_{\perp -\sigma})} + \text{H.c.}$. Due to this term, there is Josephson phase coherence between the Cooper pair condensates.  The scaling dimension associated to this is
\begin{equation}
\delta_2 = \frac{ 1 }{ 2 } + \frac{1}{2 K^1_\rho} + \frac{1}{2 K^2_\rho}.
\end{equation}

The third contribution in Eq.~(\ref{Eq:FHSG}) corresponds to the operator $(\psi_{-r,\sigma}^1)^\dagger \psi_{+r,\sigma}^2 (\psi_{-r,\sigma'}^1)^\dagger \psi_{+r,\sigma'}^2 e^{ i a' (A_{\perp \sigma} + A_{\perp \sigma'})} + \text{H.c.}$. It is a correlated hopping term that is irrelevant without longer ranged repulsive interactions
\begin{equation}
\delta_3 = \frac{1}{2} \left( K_\rho^1 + K_\rho^2 + \frac{1}{K_\rho^1} + \frac{1}{K_\rho^2} \right) > 2.
\end{equation}
We will return to this term in Subsec.~\ref{Subsec:FLa2}. It favors the Cooper pair \Laughlin ground state.

\subsection{Cooper pair \MIMe phase}
\label{Subsec:FMM}
To realize \MIMe, take $U_1 < 0$ and $U_2 > 0$ on the order of the bandwidth $4t$, such that $K_\rho^1 = 2$ and $K_\rho^2 = 1/2$. Then $\delta \equiv \delta_1 = \delta_2 = 7/4$, showing that it is possible to achieve an energy scale 
\begin{equation}
\Delta^* \sim \Delta_\sigma^1 \left(\frac{g' a}{v_F}\right)^{\frac{1}{2 - \delta}},
\end{equation}   
under which Mott insulating behavior and Meissner currents coexist. Note that the \MIMe gap lies in general below the spin gap in chain 1, \textit{i.e.} $\Delta^* < \Delta_\sigma^1$. We have let $v_F$ be a velocity close to the Fermi velocities of the two chains. Remark the difference from the bosonic case, Subsec.~\ref{Subsec:BMM}. The Cooper pair \MIMe gap has a power law dependence on the tunneling between chains. We denote this phase by \MIMe between the $\rho$ sectors of chains 1 and 2 on Figure~\ref{Fig:FermiLadder}a) .

To characterize \MIMe, let us consider the relative charge current
\begin{eqnarray}
i [ \mathcal{H}, -\frac{1}{\pi} \sqrt{2} \nabla \phi_\rho^- ] = \frac{d}{dt} (n_1 - n_2) \nonumber
\end{eqnarray}
which splits as before into two components
\begin{eqnarray}
j_\perp &=& \frac{4g'}{a} \langle \cos(  \sqrt{2}\phi_\sigma^2  )  \rangle \sin( -2\chi x + 2 \theta_\rho^- ), \\
j_\parallel &=& - v_\rho^1 K_\rho^1 \nabla \theta_\rho^1 + v_\rho^2 K_\rho^2 \nabla \theta_\rho^2 = - v_F \sqrt{2} \nabla \theta_\rho^-.
\end{eqnarray}
These operators are the analogues of Eqs.~(\ref{Eq:BMMC}). For energy scales smaller than $\Delta^*$, they have the following expectation values
\begin{eqnarray}
\langle j_\perp \rangle = 0 ,\; \langle j_\parallel \rangle = -  \sqrt{2} v_F \chi = - 2\sqrt{2} a t \chi. 
\end{eqnarray}
The new factor in the second equation comes from the fact that we are considering Cooper pairs (hence a $\sqrt{2}$) with magnetic flux $\chi$ per spin (hence the $2$).

\subsection{Cooper pair \Laughlin phase}
\label{Subsec:FLa2}
The third term in Eq.~(\ref{Eq:FHSG}) produces a Laughlin state at $\nu = \frac{1}{2}$ for the Cooper pairs. It can be made relevant by the addition of an interchain repulsive interaction.

Let us assume that the two chains are identical $K_\rho = K_\rho^1 = K_\rho^2$ and $v_\rho = v_\rho^1 = v_\rho^2$; further we assume that they have attractive interactions and that $\phi_\sigma^{1}$ and $\phi_\sigma^2$ are both gapped.

To make the third term of Eq.~(\ref{Eq:FHSG}) relevant, it is sufficient to add an interchain interaction
\begin{equation}
\mathcal{V} = \frac{a V_\perp}{\pi^2} \int dx  (\nabla \phi_\rho^1 ) (\nabla \phi_\rho^2).
\end{equation} 
We need to reexpress the Luttinger liquid Hamiltonian describing the density sector
$\mathcal{H}_{0\rho}^1 + \mathcal{H}_{0\rho}^2 + \mathcal{V} = \mathcal{H}_{0\rho}^+ + \mathcal{H}_{0\rho}^-.$ The new Luttinger liquid Hamiltonians are characterized by parameters
\begin{eqnarray}
v_\rho^\pm K_\rho^\pm = v_\rho K_\rho, v_\rho^\pm / K_\rho^\pm = \frac{v_\rho}{K_\rho} \pm \frac{a V_\perp}{\pi},
\end{eqnarray}
from which $K_\rho^\pm = (1 - u \pm v_\perp)$, where $u = |U|a/(\pi v_F)$ and $v_\perp = V_\perp a /(\pi v_F)$. Thus, we see that the scaling dimension $\delta_3 = 1/(1-u-v_\perp) + 1- (u-v_\perp) < 2$ for large enough repulsive interaction $v_\perp > 0$ between the chains.

By imposing the following constraint on the flux and density, 
\begin{equation}
\label{Eq:FLa2Cs}
2 ( k_F^1 + k_F^2 ) \pm 2 \chi  = 0 \mod 2\pi,
\end{equation}
the effective sine-Gordon Hamiltonian from Eq.~(\ref{Eq:FHSG}) is
\begin{equation}
\mathcal{H}_\perp^{II} = -\frac{g'}{a} \int dx \cos\left[ \sqrt{2} (\theta_\rho^1 - \theta_\rho^2) \pm \sqrt{2} (\phi_\rho^1 + \phi_\rho^2) \right].
\end{equation}
Upper and lower signs correspond to the constraint in Eq.~(\ref{Eq:FLa2Cs}). The canonical transformation
\begin{equation}
(1/\sqrt{2}) \Theta^\alpha_\rho =  \theta^\alpha_\rho, \sqrt{2} \Phi^\alpha_\rho = \phi^\alpha_\rho
\end{equation}
performed for each chain $\alpha$ yields the interchain coupling
\begin{equation}
\label{Eq:FLaughlinCoupling}
\mathcal{H}_\perp^{II} = -\frac{g'}{a} \int dx \cos\left[ \Theta_\rho^1 - \Theta_\rho^2 \pm 2 (\Phi_\rho^1 + \Phi_\rho^2) \right].
\end{equation}
Eq.~(\ref{Eq:FLaughlinCoupling}) is formally identical to Eq.~(\ref{Eq:LaughlinCouplingm}) and describes the Laughlin state at filling $\nu = \frac{1}{2}$ for Cooper pairs, which are created by the operator $(\psi^\alpha_{-r,\uparrow} \psi^\alpha_{r,\downarrow})^\dagger \sim e^{ i \Theta^\alpha_\rho}$. The discussion of observables in this phase is analogous to the one in Subsec.~\ref{Subsec:La2}.

\subsection{Dual phase: fermionic Mott insulator with spinon currents}
\label{Subsec:FMSp}
In this section we present a phase dual to \MIMe. \MIMe involved inducing superconductivity through the proximity effect, \SC, and the formation of \MIMe. The dual to this occurs at half-filling in each chain $k_F^1 = k_F^2 = \frac{\pi}{2a}$, when it is necessary to include the \textit{umklapp terms} in Eq.~(\ref{Eq:HFb})
\begin{equation}
\label{Eq:Umklapps}
\sum_\alpha \frac{U_\alpha}{2\pi a} \int dx \cos( \sqrt{8} \phi_\rho^\alpha ).
\end{equation}
The resulting phase will be the \MI. In this Mott phase, a spinon Meissner phase will develop, which we denote by \SMe. The phase is summarized in Figure~\ref{Fig:FermiLadder}b).

To obtain the dual phase, assume that in both chains there are repulsive interactions $U \equiv U_1 = U_2 > 0$ and that $K_\rho = K_\rho^1 = K_\rho^2$. Assume in addition that $K_\sigma^{1,2} = K_\sigma \geq 1$, which makes terms generating a spin gap in~(\ref{Eq:HFb}) irrelevant. For this, it is necessary that the repulsion $U$ be on the order of the bandwidth. The Mott gap has the asymptotic power law form,
\begin{equation}
\Delta_\rho \sim U \left( \frac{ U a }{ v_F } \right)^{1 / (2 - 2 K_\rho)}.
\end{equation} 
Then the tunneling term $\mathcal{H}_\perp^I$ is irrelevant, and we can proceed to obtain a Hamiltonian at second order in perturbation theory. The derivation of this Hamiltonian can be found in Appendix~\ref{Ap:And}. We obtain 
\begin{eqnarray}
\label{Eq:FHSG2}
\mathcal{H}_\perp^{II} = - \frac{g'}{a} \int dx \cos\left( 2 \theta_\sigma^-  -2 \chi_\sigma x \right) \langle  \cos( \sqrt{2} \phi_\rho^1 ) \cos( \sqrt{2} \phi_\rho^2 ) \rangle, \nonumber \\
\;
\end{eqnarray}
The expectation value is order 1 for energy scales under $\Delta_\rho$.  We introduced
\begin{eqnarray}
\frac{g'}{a}  &=& 4 \frac{g^2}{\Delta_\rho^1} \frac{1}{\pi a} \\
2\chi_\sigma x  &=& -a' A_{\perp\downarrow} + a' A_{\perp\uparrow} \label{Eq:chisig}.
\end{eqnarray}
For the second equation, we require that the two spin species have different charges with respect to the gauge field. This results in a flux coupling to spin, denoted $\chi_\sigma$. 

It is now easy to see that a spinon Meissner phase can arise. Under the Mott gap, the effective scaling dimension of Eq.~(\ref{Eq:FHSG2}), which represents a Josephson term for the spinon phase, is $1/K_\sigma$.
As before, we consider the low $\chi_\sigma$ limit, so the oscillatory argument of the sine-Gordon term can be neglected in the renormalization group flow equations. Therefore we introduce a new energy scale
\begin{equation}
\Delta \sim \Delta_\rho \left( \frac{g'a}{v_F} \right)^{\frac{1}{2-\frac{1}{K_\sigma}}}.
\end{equation}
This energy scale is under the Mott gap and characterizes the onset of spinon Josephson phase pinning. In analogy to the situation studied before, Meissner spinon currents are allowed in this phase 
\begin{equation}
\langle j^\sigma_\perp \rangle = 0 , \langle j^\sigma_\parallel \rangle = - 2 \sqrt{2} a t \chi_\sigma,
\end{equation}
as obtained from the time derivative of the relative spin density $-\frac{1}{\pi} \nabla \phi^-_\sigma (x)$. The $\sqrt{2}$ comes from the definition of the spinon field $(\psi^\alpha)^\dagger_{r,\uparrow} \psi^\alpha_{-r,\downarrow} \sim e^{- i 2 r k_F x} e^{-i \sqrt{2} \theta^\alpha_\sigma}$, and the factor of $2$ comes from~(\ref{Eq:chisig}).


\section{Experimental realizations}
\label{Sec:Expt}
In this section we propose experimental realizations of Eq.~(\ref{Eq:Hbt}) with ultracold atoms in optical lattices (Subsec.~\ref{Subsec:CAInt}) and with quantum circuits (Subsec.~\ref{Subsec:JJA}).

\subsection{Ultracold atom implementation}
\label{Subsec:CAInt}
\subsubsection{Ladder implementation}
Several aspects related to our model have been proven experimentally. The Abrikosov vortex lattice was observed in rotating traps \cite{DalibardVortex, *KetterleVortex}. The Josephson effect was demonstrated with spatially separated Bose-Einstein condensates \cite{KetterleJosephson,*ThywissenJosephson}. 
A recent experiment \cite{Atala} demonstrates the Meissner effect in a ladder optical lattice of about $40$ rungs filled with approximately $5 \times 10^4$ $^{87}$Rb atoms. For the purposes of this subsection, we will use the notation of Ref.~\cite{Atala} and define hopping matrix elements $J_x$ and $J_y$. The square lattice is defined by translation vectors $\mathbf{d}_{x,y}$. The tunneling along the $x$ direction can be suppressed by means of an inhomogenous electric field inducing a tilt $\Delta_\textit{tilt} \gg J_x$ between neighboring minima of the optical lattice. The tunneling can be restored resonantly by a pair of far-detuned Raman running-wave beams $(\mathbf{k}_1 , \Omega_1)$ and $(\mathbf{k}_2, \Omega_2)$. The frequency detuning $\omega = |\Omega_1 - \Omega_2|$ is matched to the tilt $\Delta_\textit{tilt}$. This driving scheme gives a spatially modulated and time-dependent potential energy at every site, $V_{m,n} = V_K^0 \cos^2\left[ \mathbf{q} . (m \mathbf{d}_x + n \mathbf{d}_y) / 2 + \omega t / 2 \right]$.  Then the wavevector $\mathbf{q} = \mathbf{k}_1 - \mathbf{k}_2$ induces a phase $\Phi_{m,n} = \mathbf{q}.(m \mathbf{d}_x + n \mathbf{d}_y)$ for hops from site $(m,n)$ to site $(m_1,n)$ (see also Ref.~\cite{Kolovsky}). The effective Hamiltonian is
\begin{equation}
H = - \sum_{m,n} \left( K e^{i \Phi_{m,n}} a^\dagger_{m+1,n} a_{m,n} + J a^\dagger_{m,n+1} a_{m,n}  \right).
\end{equation}
For large tilts $\Delta_\textit{tilt} \gg V_K^0$, the renormalized hopping strengths are $K = J_x V_K^0 / (\sqrt{2} \Delta_\textit{tilt})$, and $J \approx J_y $. The experiment for the realization of the Meissner effect in a ladder started with a finite value of the flux and increased the coupling between the wires $J$ [$g$ in the notation of our Eq.~(\ref{Eq:Hbt})] in order to obtain a Meissner phase. Decreasing $J$ allowed a transition into a vortex lattice below some critical rung hopping matrix element $J^c$. 

In the experiment of Ref.~\cite{Atala} one can, in principle, control the lattice filling such that on average an \textit{odd number of bosons per rung} is achieved. 
The charge gap of the Mott insulator can be probed by implementing an additional tilt of the ladder lattice in the $x$ direction. $\Delta^+$ can be determined from the particle-hole excitation probability in the total density sector under the tilt, as exemplified in the classic experiment by Greiner \textit{et al} \cite{Greiner2001}.

\subsubsection{Alternative implementation: Spin-orbit coupling of hyperfine states}

Let us briefly discuss an alternative implementation in the setup of Ref.~\cite{AidelsburgerHofstadter}. It is possible to formally map the chains 1 and 2 into two internal degrees of freedom of atoms in a single one-dimensional optical lattice. Coherent transport and splitting of atomic wavepackets for different Zeeman states has been demonstrated \cite{CTSpinDep}. As a concrete example, for $^{87}\text{Rb}$ atoms, the Zeeman states with opposite magnetic moments $|1\rangle = |F = 1, m_F = -1\rangle$ and $|2\rangle = |F = 2, m_F = -1\rangle$  experience opposite Peierls phases in the presence of laser assisted tunneling \cite{AidelsburgerHofstadter}.  This amounts to a spin-orbit coupling term for the spinor Bose gas in one dimension.

The Josephson term couples the two Zeeman states, taking the form 
\begin{equation}
-g | 1 \rangle_i \langle 2 |_i - g | 2 \rangle_i \langle 1 |_i \nonumber 
\end{equation} 
for an atom at site $i$. The Peierls phase corresponds to the transport of an atom from site $i$ to site $i+1$, $-t e^{i a A^\alpha_{i,i+1} }| \alpha \rangle_i \langle \alpha |_{i+1} + \text{H.c.}$, where $\alpha = 1,2$ and $A^{1}_{i,i+1} = - A^2_{i,i+1}$. Importantly, the odd integer filling condition~(\ref{Eq:TotalDensity}) becomes a simple condition on the \textit{parity of the atom number at each site}:
\begin{equation}
\big\langle \, |1\rangle_i \langle 1|_i + |2\rangle_i \langle 2|_i \, \big\rangle \equiv 1 \text{ mod }2.
\end{equation} 
The odd-integer filled \MI can then be probed since atom number parity can be imaged with current technology \cite{Bakr2009,Sherson2010}. The \MIMe can be obtained by preparing an odd filling Mott insulator, then tuning the population of  $|1\rangle$ versus $|2\rangle$ by the application of microwave fields and a magnetic field to realize a Landau Zener sweep \cite{AidelsburgerHofstadter}.

The \Laughlin phase becomes possible for a finite value of $V_\perp > 0$. Interactions between distinct spin species \cite{StamperKurnUeda2013} can be tuned by magnetic Feshbach resonances \cite{ChinFeshbach}, which is a possible pathway towards stabilizing the \Laughlin phase. In the \Laughlin phase, a lattice tilt would yield spin flip current.

More generally, long ranged repulsion between next-neighbor atoms can be achieved with the dipole-dipole interactions of Rydberg atoms \cite{Saffman,RydDC}. Dipolar molecule interactions can be used as well \cite{Dip}. We have argued previously that free fermions interacting repulsively with the bosons give rise to effective repulsive interactions between the bosons \cite{1}.

\subsection{Quantum circuit implementation}
\label{Subsec:JJA}
\begin{figure}[t!]
\includegraphics[width=1.00\linewidth]{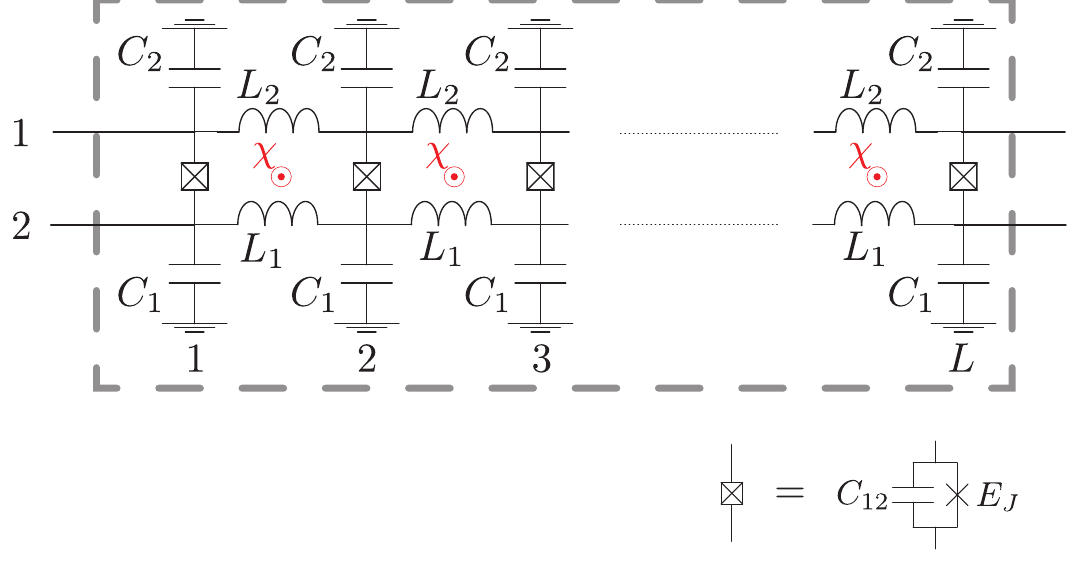}
\caption{\label{Fig:JJA}Two-leg ladder as quantum circuit, with Josephson junction components on the rungs.}
\end{figure}

\begin{figure}[t!]
\includegraphics[width=0.80\linewidth]{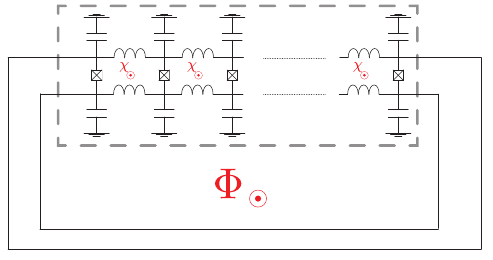}
\caption{\label{Fig:JJAAB} Setup that allows threading of Aharonov Bohm flux, denoted $\Phi$, through the ladder.}
\end{figure}

It has long been known that the \Me and \VL phases can be realized in Josephson junction arrays \cite{Kardar1986, Granato1990, DennistonTang1995}. When the Josephson coupling energy is comparable to the capacitive charging energy of the superconducting islands, such systems present a magnetic field tuned superconductor -- insulator transition \cite{Chen1995,Zant1996}. At small Josephson energy, the dynamics of the system is described in terms of charges, whereas the large Josephson energy limit gives way to a description in terms of vortices. The field tuned superconductor-insulator transition is a vortex delocalization transition, proposed by Fisher \cite{Fisher1990}.  Moreover, quantum Hall states have been theoretically predicted for Josephson junction arrays: quantum Hall phases of vortices stabilized by inherent long ranged interactions \cite{Choi1994,Stern1994}; and, directly relevant to our discussion, a fractional quantum Hall state at $\nu = \frac{1}{2}$ was predicted in Josephson arrays by Odintsov and Nazarov \cite{OdintsovNazarov1994}.

We propose here a quantum circuit \cite{Devoret,DevoretMartinis} realization of the Hamiltonian in Eq.~(\ref{Eq:Hbt}). In this circuit, the various energy scales are tuned to agree with the effective continuum theory~(\ref{Eq:HWB}). We associate to the capacitive, inductive and Josephson junction circuit components energy scales
\begin{eqnarray}
E_C = \frac{e^2}{2C}, \; E_J = \frac{I_c}{\phi_0}, \; E_L = \frac{\phi_0^2}{L}.  
\end{eqnarray}
Here $\phi_0 = \hbar/(2e)$ is the reduced flux quantum. The critical current $I_c$ of the Josephson junction is in the nA-$\mu$A range. Typical capacitances $C$ are in the fF to pF range. Inductances $L$ can be in the nH range.

To define the quantum circuit Hamiltonian, we define node fluxes \cite{Devoret} $\theta_{i}^{1,2}$ on the circuit in Figure~\ref{Fig:JJA}. Josephson junctions connecting the chains 1 and 2 correspond to cosine terms in the circuit Hamiltonian
\begin{equation}
\label{Eq:HJ1}
-\sum_{i=1}^L E_J^{12} \cos( \theta_{i}^1 - \theta_{i}^2 + a' A_{\perp i} ).
\end{equation}
In addition to this, the mutual capacitance $C_{12}$ between the chains leads to a charging term
\begin{equation}
\label{Eq:HJ2}
\sum_{i=1}^L E^{12}_C (n_{1i} - n_{1i}^0 )( n_{2i} - n_{2i}^0 ).
\end{equation}
The offset charges on each superconducting island are denoted $n_{\alpha i}^0$.

We turn now to terms corresponding to individual chains. There is a charging energy at the $i^\textit{th}$ site in chain $\alpha$ due to the capacitive coupling $C_\alpha$ 
\begin{equation}
\label{Eq:HJ3}
\sum_{i=1}^L E_C^\alpha (n_{\alpha i} - n_{\alpha i}^0)^2.
\end{equation}
Additionally, we have assumed that the Josephson energy associated with a junction between sites $i$ and $i+1$ is large compared to the charging energy $E_J^\alpha \gg E_C^\alpha$. Then Josephson terms in each chain are replaced by inductive contributions
\begin{equation}
\label{Eq:HJ4}
\sum_{i=1}^{L-1}\frac{E_J^{\alpha}}{2} \left(\theta^{\alpha}_i - \theta^{\alpha}_{i+1} + a A_{i,i+1}^\alpha \right)^2.
\end{equation} 

The Hamiltonian is the sum of Eqs.~(\ref{Eq:HJ1}, \ref{Eq:HJ2}, \ref{Eq:HJ3}, \ref{Eq:HJ4}). We now estimate the involved energy scales. The chain Josephson energy scales must be set large $E_J^{1,2} / h \approx 10$\;GHz, compared to charging energy  $E_C^{1,2} / h \approx E_J^{12} / h \approx  2 $\;GHz. These values are commonly achieved in experiments \cite{NMaslukThesis, Bibow}. Note that typical temperatures are 20 mK corresponding to frequencies of 0.4 GHz. This is well below the superconducting gap of aluminum, about 2 K. Returning to the notation of the original Hamiltonian in Eq. (\ref{Eq:Hbt}),
\begin{equation}
t \sim E_{J}^{\alpha},\;U = E_C^\alpha, \; g = E_J^{12},\; V_\perp = E_{C}^{12}.
\end{equation}
The Luttinger parameter in each chain is very large if $E_C^{\alpha}$ is negligible and the Tonks limit would be achieved in a limit where the intra-chain charging energy would formally become infinite. 

One shared characteristic of Josephson junction array experiments is the presence of offset charge noise \cite{DevoretMartinis}, which becomes difficult to control over large arrays \cite{Zant1996}.  Control of offset charge is crucial for the realization of the \MI phase. Offset charges can in principle be tuned by voltage terms $- V_i^{\alpha} (2e) n_{\alpha i}$. With the aid of these terms and tuning the mutual capacitance $C_{12}$, it is possible to achieve a stable state with an odd number of Cooper pairs on each rung. This was shown in the context of a pair of superconducting islands \cite{Bibow,JensKarynJJ}. While control of offset charge over a large array is hard experimentally, signatures of the phases proposed here should in principle appear in arrays of \textit{several} junctions. The charge gap $\Delta^+$ can be probed by showing the absence of current when flux is threaded through the cylinder of the ladder, as argued in Sec.~\ref{Subsubsec:LaughlinArgMJ}. Experimentally this is achieved by threading flux through two large external loops that wrap around the cylinder (Figure~\ref{Fig:JJAAB}).  In the \Laughlin phase, current through the rungs should be observed while adiabatically threading AB flux through the ladder.


\section{Two-dimensional generalizations}
\label{Sec:Nleg}

In this section we generalize the 2-chain ladder models of Sec.~\ref{Sec:WB} to $N$-leg ladders. We find that the \MIMe cannot be stable for $N>2$ if we keep the same average filling per chain is $n_0 = \frac{1}{2a}$. However, regardless of filling, Josephson phase pinning terms are present and allow us to generalize the \Me phase in $N$-leg ladders. This is detailed in Sec.~\ref{Subsec:NchainMe}. We find that the \MIMe can be actually stabilized if the ladder has $(N-1)$ bosons per unit cell. In this case, a Mott phase can be stabilized on the inner chains of the ladder, whereas \MIMe occurs on the outer chains. This is presented in Sec.~\ref{Subsec:3leg}. Finally, we dedicate Subsec.~\ref{Subsec:2pla} to two-dimensional generalizations which involve bilayers formed by juxtaposing ladders.

\subsection{$N$-chain construction for the Meissner phase}
\label{Subsec:NchainMe}
\begin{figure}[t!]
\includegraphics[width = 0.90 \linewidth]{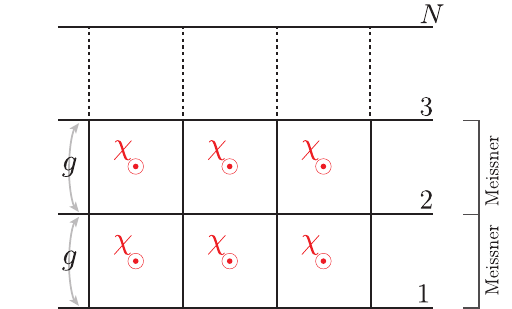}
\caption{\label{Fig:Nleg}Planar array of $N$ chains obeying Eq.~(\ref{Eq:H2DBM}). In the strong coupling phase, only edge bonds in chains $1$ and $N$ carry nonvanishing current.}
\end{figure}

\begin{figure}[t!]
\includegraphics[width=0.90\linewidth]{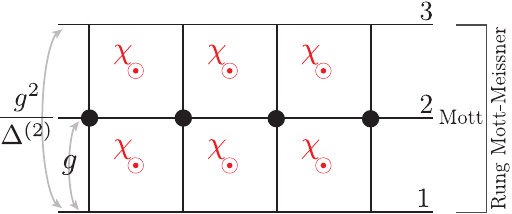}
\caption{\label{Fig:3leg} Three leg ladder construction for \MIMe. There is Mott insulator of average filling $n_0^2 = 1/a$ on wire 2. The Josephson coupling between chains $1$ and $3$ appears at second order in perturbation theory.}
\end{figure}

Consider $N$ identical bosonic chains (Figure~\ref{Fig:Nleg}) described by Luttinger liquid Hamiltonians as in Eq.~(\ref{Eq:LLp},\ref{Eq:LLm}): the fields corresponding to the $J^\textit{th}$ chain are $\theta^J(x)$ and $\phi^J(x)$. Their commutation relation is $[ \phi^{J}(x), \theta^{J'}(x') ] = i \frac{\pi}{2} \delta_{JJ'} \text{Sign}(x'-x)$. Under the assumptions of Subsec. \ref{Subsec:BMM} for density and flux, we obtain the following coupling Hamiltonian
\begin{eqnarray}
\label{Eq:H2DBM}
\mathcal{H}^{\text{2D}}_\perp &=& \sum_{J=1}^{N-1} \mathcal{H}^{J,J+1}_\perp \nonumber \\
\mathcal{H}^{J,J+1}_\perp &=& -\frac{g}{a} \int dx \cos( \theta^{J+1} - \theta^{J} - \chi x ) \times \nonumber \\ && \;\;\;\;\;\;\;\;\;\;\;\;\;\;\;\left(1 +  2\cos\left[ 2 (\phi^J + \phi^{J+1} )\right]\right) .
\end{eqnarray}
Under the energy scale $\Delta^-$ of Eq.~(\ref{Eq:Deltam}), the phase fields $\theta^{J+1} - \theta^J$ are pinned to the classical value $\langle \theta^{J+1} - \theta^J \rangle = - \chi x$. The following Meissner currents result
\begin{eqnarray}
\label{Eq:EVj}
j_\perp &=& 0, \nonumber \\
j_{\parallel}^J &=& 0, \; J = 2,...,N-1  \nonumber \\
j_\parallel^{1} - j_\parallel^N &=& - ( N - 1 ) a t \chi.
\end{eqnarray}

Concerning the possibility of a Mott transition, the $N-1$ fields $\phi^{1} + \phi^{2}$, ..., $\phi^{N-1}+\phi^N$ cannot be pinned, since they are not independent from the fields $\theta^{1} - \theta^{2}$, ..., $\theta^{N-1}-\theta^{N}$. This is because there is a nonzero commutation relation:
\begin{eqnarray}
[ \phi^{J}(x) + \phi^{J+1}(x), \theta^{J}(x') - \theta^{J - 1}(x') ] &=& [ \phi^{J}(x), \theta^{J}(x') ] \nonumber \\
&=& i \frac{\pi}{2} \text{Sign}(x'-x). \nonumber
\end{eqnarray}
The field $\phi^{\rho} = \frac{1}{\sqrt{N}} \sum_{j=1}^{N} \phi^j$ describing fluctuations of the total density remains gapless, leading to a power law density-density correlation function. In conclusion, the $N$-leg ladder leads to a Meissner effect with vanishing bulk current expectation values. The Mott phase is unstable for $N > 2$. The Mott phase described by a finite correlation length discussed for $N=2$ is replaced by  algebraic density-density correlation functions for $N\geq3$.

If the flux and density are changed to satisfy $\nu = \frac{1}{2}$ filling, the coupled chain construction of the bosonic Laughlin state is obtained.  For $N \geq 3$, it is possible to form a closed loop contained entirely in the bulk, at each point on the loop having one vertical bond (rung). Fractional statistics are manifest in the phase acquired by bulk quasiparticles around such a closed loop \cite{TeoKane}.

\subsection{3-leg construction for the boson Mott insulator with Meissner current}
\label{Subsec:3leg}
Let us return to the possibility of realizing \MIMe in an $N$-leg ladder. In order to achieve this phase, the densities need to be changed from half-filling. In the simplest instance, it is possible to realize a Mott insulator with Meissner current at the edges in a 3-leg ladder (see Fig.~\ref{Fig:3leg}). Consider the Hamiltonian of Eq.~(\ref{Eq:Hbt}), with $\alpha = 1,2,3$ denoting the three chains. Let $n_0^2 = \frac{1}{a} = n_0^{1} + n_0^3$. Then the bosonized form of the chain number 2 is:
\begin{eqnarray}
\mathcal{H}^2 &=& \frac{v^2}{2\pi} \int dx \left[  K^2 (\nabla \theta^2)^2 + \frac{1}{K^2}(\nabla \phi^2)^2 \right] \nonumber \\
&& + \frac{U_2}{a} \int dx \cos\left( 2 \phi^2 \right).
\end{eqnarray}
Assuming $K^2 < 2$ and $U_2 > U_{1,3}$, the middle chain undergoes a Mott transition characterized by the energy scale
\begin{equation}
\Delta^{(2)} \sim \frac{v^2}{a} \left( \frac{a U_2}{v^2} \right)^{\frac{1}{2 - K^2}}.
\end{equation}
The Josephson coupling terms are analogous to Eq.~(\ref{Eq:HSG}). For $T < \Delta^{(2)}$, where $\theta^2$ is disordered, the effective Josephson coupling between chains 1 and 3 appears at order $\frac{g^2}{\Delta^{(2)}}$:
\begin{widetext}
\begin{eqnarray}
\mathcal{H}_{\textit{SG}} &=& -\frac{4 g^2}{a \Delta^{(2)}} \int dx \cos(-\theta^1 + \theta^2 + \chi x) \cos( -\theta^2 + \theta^3 + \chi x) \left[ 1 + 2\cos\left( 2\pi  n_0^1 x - 2 \phi^1 \right)  \right] \, \left[ 1 + 2\cos\left( 2\pi  n_0^3 x - 2 \phi^3 \right)  \right] \nonumber \\
&=& - \frac{2 g^2}{a \Delta^{(2)}} \int dx \cos( - \theta^1 + \theta^3 + 2 \chi x ) \, \left[ 1  + 2 \cos\left( 2 \phi^1 + 2 \phi^3 \right) \right] + ...
\end{eqnarray}
The ellipsis contains a term proportional to $\cos(-\theta^1-\theta^3+2\theta^2)$, which is less relevant. The remaining contribution is identical to Eq.~(\ref{Eq:HSG}), provided that the following changes are made: field $\phi^2$ becomes $\phi^3$, $\theta^2 \to \theta^3$, and $\chi \to 2\chi$. The transport observables are obtained from Subsec.~\ref{Subsec:BMM} with the substitutions described in this paragraph.
\end{widetext}

The description of the phase for temperatures below $\Delta^{(2)}$ is analogous to that of the two chain ladder with double the flux, and a suppressed Josephson coupling $\propto g^2/\Delta^{(2)}$.  The Josephson phase pinning gap $\Delta^-$ is modified to account for the fact that the renormalization group flow begins at the Mott energy scale of the middle chain, $\Delta^{(2)}$:
\begin{equation}
\Delta^- = \Delta^{(2)} \left( \frac{g^2}{\Delta^{(2)}} \frac{a}{v} \right)^{\left( 2-\frac{1}{2 K^-} \right)^{-1}},
\end{equation}
with $K^-$ as given in Subsec.~\ref{Subsec:BMM} in Eq.~(\ref{Eq:vKWB}). From here,  $\Delta^+$ is obtained via Eq.~(\ref{Eq:Deltap}). Then we expect the following hierarchy 
\begin{equation}
\Delta^+ < \Delta^- < \Delta^{(2)}.
\end{equation}
The large Mott gap in the middle chain, $\Delta^{(2)}$, implies that an added particle will go to one of the outer chains $1$ or $3$. This causes a \MI to transition to \SF. Therefore doping leads to the phase \SFMe, along with a Mott insulator at unit filling in chain 2. This situation is reminiscent of the d-Mott phase of spinful fermion ladders, where a hierarchy of gaps leads to similar behavior \cite{Lin1997,*Ledermann2000,*Karyn2009}.

The presence of the unit filled Mott insulating chain $2$ induces Josephson coupling between chains $1$ and $3$ at order $g^2/\Delta^{(2)}$. To generalize, assume that the ladder consisted of $N+2$ chains, $N$ of which were at unit filling in a Mott phase, making up the bulk. The Josephson term between chains $1$ and $N+2$ would appear at order $g^{N+1}/(\Delta^{(2)})^N$ in perturbation theory. This is the exponential suppression of the tunneling term between the edge chains $1$ and $N+2$ due to the insulating bulk.

\subsection{2D construction for fermionic Mott insulator with spinon currents}
\begin{figure}
\includegraphics[width=0.90\linewidth]{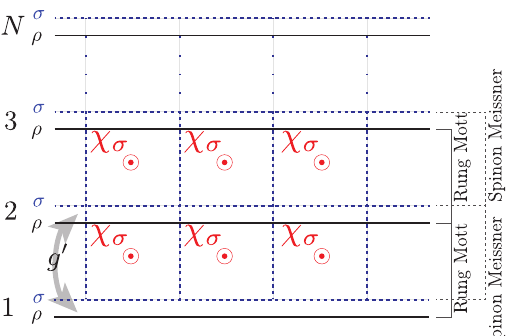}
\caption{\label{Fig:MIpSMem} Two dimensional \MISMe, obtained from Eq.~(\ref{Eq:MIpSMem})}
\end{figure}

The coupling Hamiltonian of Eq.~(\ref{Eq:FHSG2}) can be generalized to $N$ identical chains [see Fig.~\ref{Fig:MIpSMem}]
\begin{eqnarray}
\label{Eq:MIpSMem}
\mathcal{H}^{2D}_\perp &=& \sum_J \mathcal{H}^{J,J+1}_\perp  \\
\mathcal{H}^{J,J+1}_\perp &=& -\frac{g'}{a} \int dx \cos\left( \sqrt{2} \phi^J_\rho \right) \cos \left( \sqrt{2}\phi^{J+1}_\rho \right) \times \nonumber \\ &&\;\;\;\;\;\;\;\;\;\;\;\cos\left[ - 2 \chi_\sigma x + \sqrt{2} ( \theta^J_\sigma - \theta^{J+1}_\sigma )  \right] \nonumber . 
\end{eqnarray}
Here $1 \leq J \leq N-1$, and $\mathcal{H}^{J,J+1}_\perp$ is the same as $\mathcal{H}^{II}_\perp$ of Eq.~(\ref{Eq:FHSG2}) only for fields corresponding to chains $J$ and $J+1$ instead of $1$ and $2$. The scaling dimension of the sine-Gordon operator is the one calculated in Subsec.~\ref{Subsec:FMSp}. Assuming that only the field $\phi_\rho^1$ is pinned to its classical value, the sine-Gordon terms pin the density fields $\phi_\rho^J$, for all remaining chains $J \geq 2$, inducing a Mott transition in each chain. In addition, terms dependent on the spinon field phase differences $\theta_\sigma^J - \theta_\sigma^{J+1}$ cause a Meissner effect. Current vanishes on all ``bulk'' chains $2 \leq J \leq N-1$ and on all vertical bonds (between chains 1 and 2,..., $N-1$ and $N$). The Meissner current at the edge is $\langle j^\sigma_N - j^\sigma_1  \rangle = - 2\sqrt{2} (N-1) a t \chi_\sigma$, as found in Sec.~\ref{Subsec:FMSp} for $N=2$.

\subsection{Coupled planes}
\label{Subsec:2pla}
\begin{figure}
\includegraphics[width=0.90\linewidth]{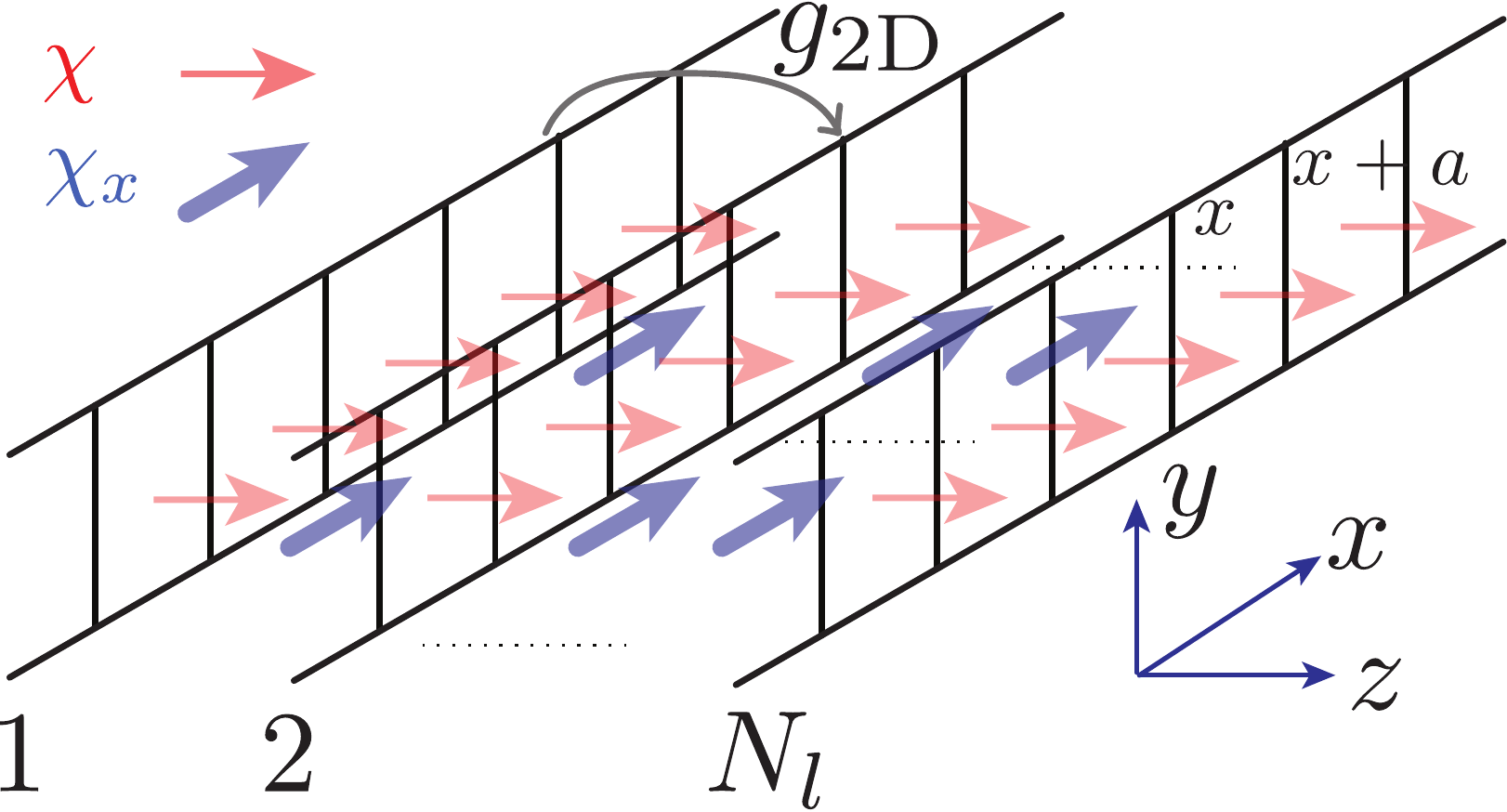}
\caption{\label{Fig:2L}Bilayer formed by an array of $N_l$ two leg ladders. The $\chi$ flux [see Eq.~(\ref{Eq:HSG})] is produced by a magnetic field in the $z$ direction (thin red arrows). A magnetic field in the $x$ direction (thick blue arrows) has flux $\chi_x$ through the vertical square plaquettes lying parallel to the $yz$ plane. }
\end{figure}
It is possible to generalize the models of Sections~\ref{Sec:WB} and ~\ref{Sec:WF}  by Josephson coupling more ladders to form bilayers.  In the bilayer geometry, different magnetic field orientations highlight the Meissner effect. The two-dimensional extension of \MIMe is a stack of ladders in the \MIMe phase; nonetheless, the ground state has spinon superfluidity and Meissner currents between any two consecutive ladders. The two-dimensional extension of \Laughlin is a stack of ladders in the \Laughlin state; however, a nontrivial feature of the bilayer ground state is that the charge fluctuations of the ladders are pinned to each other.

We start with an array of two chain ladders such that the $\alpha =1,2$ wire of the $I^{\textit{th}}$ ladder lies in the $\alpha^\textit{th}$ plane of the bilayer (Figure~\ref{Fig:2L}).  Let the fields in the $I^\textit{th}$ ladder be $\theta^{1I},\phi^{1I},\theta^{2I},\phi^{2I}$. We replace $\theta^{1,2},\phi^{1,2}$ by these fields in Eqs.~(\ref{Eq:HWB})-(\ref{Eq:LLm}) and (\ref{Eq:HSG}) (\MIMe) or (\ref{Eq:LaughlinCouplingm}) (\Laughlin) to obtain the Hamiltonian for the $I^\textit{th}$ ladder. In addition, we assume that the Luttinger parameter $K$ and the velocity of excitations $v$ are independent of the ladder index $I$.

Let us assume that hopping terms across ladders, which are characterized by the hopping strength $g_{\text{2D}}$ along the $z$ direction, contain a Peierls phase $a A^\alpha_{I,I+1}$. This is given by a magnetic field parallel to the $x$ direction in Figure~\ref{Fig:2L}. We will pick a gauge such that the inter-ladder coupling is
\begin{equation}
\label{Eq:H2DB}
\mathcal{H}_{\text{2D}} = -\frac{g_{\text{2D}}}{a} \sum_{\alpha=1,2} \sum_{I=1}^{N_l-1} \int dx \cos(\theta^{\alpha I} - \theta^{\alpha,I+1} - a A^{\alpha}_{I,I+1}),
\end{equation}
where $N_l$ is the number of ladders forming the planar bilayer. Eq.~(\ref{Eq:H2DB}) is the Josephson coupling within each layer of our bilayer. The flux through a plaquette parallel to the $yz$ plane is
\begin{equation}
\chi_x = - a A^1_{I,I+1} + a A^2_{I,I+1}.
\end{equation}
To study the relevance of the hopping term in Eq.~(\ref{Eq:H2DB}), we rewrite the sum over $\alpha$ in terms of $\theta^{-,I}$ and $\theta^{+,I}$
\begin{eqnarray}
\label{Eq:2DSG}
\mathcal{H}_{\text{2D}} = -2\frac{g_{\text{2D}}}{a} \sum_{I=1}^{N_l-1} \int dx 
\cos\left(\frac{\theta^{-,I} - \theta^{-,I+1}}{\sqrt{2}} + \frac{\chi_x}{2} \right) \times \nonumber \\ 
\cos\left( \frac{\theta^{+,I} - \theta^{+,I+1}}{\sqrt{2}} - \frac{a A_{I,I+1}^1 + a A_{I,I+1}}{2} \right) \nonumber. \\ \;
\end{eqnarray}

\subsubsection{Phase locking in an array of ladders}
If the identical ladders are either in state \MIMe or \Laughlin, the term (\ref{Eq:2DSG}) is irrelevant due to the second cosine factor of the integrand. We can check this by inspecting the ordered fields in either of the two phases [described by~(\ref{Eq:HSGMM}) or (\ref{Eq:LaughlinCouplingm})], and using the fact that $[\phi^{+,I},\theta^{+,I}] \neq 0$. Nonetheless a contribution at second order in perturbation theory is relevant (see Appendix~\ref{Ap:SecondOrderPT}). To derive this contribution we need to determine the relevant energy gap that determines the correlation length. This gap is $\Delta^+$ for \MIMe [see Eq.~(\ref{Eq:Deltap})] or $\Delta$ for \Laughlin [see Eq.~(\ref{Eq:DelLa})]. For generality we denote the gap by $\Delta^{\textit{ladder}}$. Unless mentioned otherwise, the results below hold regardless of the phase of the individual ladders.

The effective Hamiltonian at second order in perturbation theory in $g_{\text{2D}}$ is:
\begin{equation}
\label{Eq:EffectiveJosephson}
\mathcal{H}_{\text{2D}} = -\frac{g^2_{\text{2D}}}{a \Delta^{\textit{ladder}}} \int dx \cos( \sqrt{2} \theta^{-,I} - \sqrt{2} \theta^{-,I+1} + \chi_x ).
\end{equation}
We have obtained a planar Meissner phase formally equivalent to the one produced by the Josephson term in Eq.~(\ref{Eq:H2DBM}). Note however that the bosons have been replaced by (pseudo-)spinons $b_{2I}^\dagger b_{1I} \propto e^{- i \sqrt{2} \theta^{-,I}}$. The spinon corresponds to interwire hopping inside the $I^\textit{th}$ ladder. The oscillatory argument of the cosine [$\chi x$ in Eq.~(\ref{Eq:H2DBM})] is absent in Eq.~(\ref{Eq:EffectiveJosephson}). Therefore, the spinon Hamiltonian has no effective magnetic field.  With this, the phase pinning condition Eq.~(\ref{Eq:EffectiveJosephson}) implies that
\begin{equation}
\label{Eq:ThetaMinIJPinning}
\langle  \theta^{-,I} \rangle = \langle \theta^{-,I+1} \rangle - \frac{1}{\sqrt{2}}\chi_x . 
\end{equation}
Using Eq.~(\ref{Eq:BMMC}) for current operators in the $I^{\textit{th}}$ ladder, we obtain
\begin{equation}
j_\parallel^I = j_\parallel^{1,I} - j_\parallel^{2,I} =  - v^- K^-  \sqrt{2} \nabla \theta^{-,I}.
\end{equation}
Then, using Eq.~(\ref{Eq:ThetaMinIJPinning}), we find that expectation values of currents in adjacent ladders obey
\begin{equation}
\label{Eq:jjLock}
\langle j_\parallel^I \rangle = \langle j_\parallel^{I+1} \rangle.
\end{equation}
The current flows are pinned to each other. 

\subsubsection{Drag effects in bilayer geometry}
We obtain here a Hamiltonian that manifests drag between Luttinger liquids \cite{KlesseStern,FieteLeHurBalents, Flensberg} due to the Josephson coupling in Eq.~(\ref{Eq:H2DB}). Start with many ladders coupled with Eq.~(\ref{Eq:H2DB}). Assume that there are Laughlin terms of the form~(\ref{Eq:LaughlinCouplingm}) in each ladder:
\begin{equation}
\label{Eq:LCI}
- 2 g n_0 \int dx \cos\left[ - (\theta^{I1} - \theta^{I2})  + m( \phi^{I1} + \phi^{I2} ) \right].
\end{equation}
Assume, as opposed to the previous subsection, that the intralayer coupling Eq.~(\ref{Eq:H2DB}) is the most relevant term. Eq.~(\ref{Eq:H2DB}) realizes the phase pinning 
\begin{equation}
\langle \theta^{\alpha I} - \theta^{\alpha,I+1} \rangle - a A^{\alpha}_{I,I+1} = 0,\; \alpha = 1, 2.
\end{equation} 
Then the Laughlin terms (\ref{Eq:LCI}) are irrelevant at first order in perturbation theory but give a contribution at second order:
\begin{equation}
- \cos( - m \sqrt{2} \phi^{+,I} + m\sqrt{2} \phi^{+,I+1} ).
\end{equation}
This term could have interesting consequences for transport.
For example, this is the form studied in the context of Coulomb drag between two (electron-like) Luttinger liquids [Eq.~(4) of Ref.~\cite{KlesseStern}]. In that case, drag resistivity is defined as $\rho_{I,I+1,d} = - \frac{V^{I+1}}{j^I L}$, where ladder $I$ is the active channel, where a drive current is applied, and ladder $I+1$ is the passive channel (not connected to any reservoirs) where a voltage $V^{I+1}$ is measured. The results known from that problem \cite{KlesseStern, FieteLeHurBalents, Pustilinik2003, Flensberg} can be used here to describe the response at finite temperature.  Importantly, in our setup, the drag is not necessarily brought about by Coulomb interactions between the ladders, but by the intralayer Josephson coupling Eq.~(\ref{Eq:H2DBM}). Drag responses have been recently measured in pairs of quantum wires separated by a small barrier \cite{Laroche2014}. Transport in these bosonic analogues will be studied elsewhere. There has been as well recent work on topologically ordered states of bosons in bilayers \cite{Repellin2014,*Moeller2014}.


\section{Conclusions}
\label{Sec:Conc}
To conclude, we have presented tight-binding models of bosons and fermions on quasi one-dimensional lattices whose ground states are incompressible states with correlations illustrating chiral order. In particular, we have shown that a Mott insulating phase with pseudospin Meissner effect \cite{1} can be observed in current experiments with ultracold atoms and Josephson junction arrays, and provided concrete experimental proposals for both setups. Moreover, we have argued that in the presence of repulsive interactions this phase transitions into  a low dimensional precursor of the Laughlin state, and enumerated observables that can be used to detect this transition. The model presented here is not restricted to bosons, and we have argued that a larger variety of phases can be obtained in a spinful fermion ladder at or near half-filling. Finally, we have derived extensions of the phases found on the ladder to two-dimensional lattices, either single-layered or bilayers. 

We remark that a recent proposal appeared for the realization of the infinitely thin cylinder limit of the Laughlin state  \cite{Grusdt2014}. The filling factor and the gauge field configuration are distinct from ours. Ref. \cite{Piraud2014} contains an extensive DMRG treatment complemented by bosonization which yields the phase diagram as a function of boson filling factor, on-site interaction $U$ and flux; where regimes coincide, our results agree.

\section*{Acknowledgements}
It is a pleasure to thank I. Bloch, J. Esteve, M. Freedman, J. Gabelli, T. Giamarchi, F. Grusdt, M. Piraud, I. M. Pop, T. M. Rice, G. Roux, and A. Tokuno for discussions. We thank Ronny Thomale for pointing out a connection to the spin liquid state proposed by Kalmeyer and Laughlin. This work has also benefitted from discussions at the CIFAR meetings on Quantum Materials in Canada, the 2014 Les Houches Summer School on Topological Aspects in Condensed Matter Physics, and the Journ\'{e}es de la Mati\`{e}re Condens\'{e}e in Paris.


\appendix

\section{Renormalization group equations for sine-Gordon models}
\label{Ap:RGE}
We consider a generic sine-Gordon Hamiltonian
\begin{equation}
H = \frac{1}{2\pi}\int dx \left( u K (\nabla \theta)^2 + \frac{u}{K}(\nabla \phi)^2  \right) + \frac{g}{a} \int dx \cos\left( \beta \phi \right),
\end{equation}
for which we derive the renormalization group equations for $g$ and $K$ to second order in the perturbation $g$. We will finally use a duality relation to derive the renormalization-group equations for $g\int dx \cos(\beta \theta(x))$. The dimensionless $\beta$ is related to the scaling dimension via $\delta = \beta^2/4$. 

Following Ref. \cite{Giamarchi}, we require that the two-point correlation function remain invariant under a change of the low distance cutoff. We expand the interacting theory zero temperature two-point correlation function
\begin{equation}
R(r_1 - r_2) = \langle e^{i \phi(r_1)} e^{- i  \phi(r_2)}\rangle
\end{equation}
to second order in $g$. 

\begin{widetext}

The expansion of the correlation function (equivalently, of the partition function) to second order in the coupling $g$ is
\begin{eqnarray}
\label{Eq:expansion_2nd_order}
R(r_1 - r_2) =  \langle e^{i \phi_1} e^{- i  \phi_2}\rangle_0 
+ \frac{1}{2^3} \left( \frac{g}{u a} \right)^2 \sum_{\epsilon',\epsilon''=\pm 1} \int d^2 r' d^2 r'' \langle e^{i\phi_1} e^{-i \phi_2} e^{i \epsilon' \beta \phi'} e^{-i \epsilon'' \beta \phi''}  \rangle_{0,\text{c}} 
\end{eqnarray}

For brevity, we denote $\phi_1 = \phi(r_1)$ etc. Integrals $\int d^2 r \equiv u \int_0^\infty dx \int_0^\infty d\tau$. The connected correlation function means $\langle e^{i\phi_1} e^{-i \phi_2} e^{i \epsilon' \beta \phi'} e^{-i \epsilon'' \beta \phi''}  \rangle_0 -\langle e^{i\phi_1} e^{-i \phi_2} \rangle_0 \langle e^{i \epsilon' \beta \phi'} e^{-i \epsilon'' \beta \phi''}  \rangle_{0}$.

In the gaussian theory correlation functions of products of exponentials are power-laws:
\begin{equation}
\left\langle \prod_j e^{i A_j \phi_j} \right\rangle_0 = \delta\left(\sum_j A_j\right) e^{-\frac{K}{2} \sum_{i < j} A_i A_j F(r_i - r_j)},
\end{equation}
where $F(r_i - r_j) \equiv \log |r_1-r_2|/a$ and the length $a$ is the small distance cutoff. The simplest of these correlation functions is the two-point correlation function $R_0(r_1-r_2) = \left( a / |r_1 -r_2|\right)^{\frac{K}{2}}.$

The double integral in Eq. (\ref{Eq:expansion_2nd_order}) is dominated by contributions from nearby terms $r' \approx r''$. Expanding in the small parameter $r = r' - r''$, we arrive at:

\begin{eqnarray}
R(r_1-r_2) = R_0(r_1 - r_2 )\Big( 1+ \frac{y^2 \beta^2 K^2}{2^5} F(r_1-r_2) \int_{r>a}\frac{dr}{a} \left(\frac{r}{a} \right)^{3-\frac{\beta^2}{2}K} \Big). \nn
\end{eqnarray}
We have introduced the dimensionless coupling constant $y = \frac{g a}{ u }$. Approximating the parenthesis by an exponential function yields
\begin{equation}
R(r_1 - r_2) \approx e^{-\frac{K}{2} F( r_1 - r_2)} e^{\frac{y^2 \beta^2 K^2}{2^5} F(r_1-r_2) \int_{r>a}\frac{dr}{a} \left(\frac{r}{a} \right)^{3-\frac{\beta^2}{2}K}}.
\end{equation} 
\end{widetext}
We express the two-point correlator as 
\begin{equation}
R(r_1 - r_2) = e^{-\frac{K_\textit{eff}}{2} F( r_1 - r_2)}
\end{equation}
with
\begin{equation}
K_{\textit{eff}}(a)  =  K - \frac{\beta^2 y^2 K^2}{2^4}\int_a^\infty \frac{dr}{a}\left( \frac{r}{a} \right)^{3-\frac{\beta^2}{2} K}.
\end{equation}
The renormalization-group equations arise by requiring that $R(r_1-r_2)$, or equivalently $K_\textit{eff}$, be invariant under a change of the low distance cutoff. We may rewrite the equation above as
\begin{eqnarray}
K_{\textit{eff}}(a) = K -  \frac{\beta^2 y^2 K^2}{2^4} \left( \int_a^{a+da} + \int_{a+da}^\infty \right) \frac{dr}{a}\left( \frac{r}{a} \right)^{3-\frac{\beta^2}{2} K}  \nn \\ 
= K - \frac{\beta^2 y^2 K^2}{2^4} \frac{da}{a} -\frac{\beta^2 y^2 K^2 }{ 2^4 } \int_{a+da}^\infty \frac{dr}{a}\left( \frac{r}{a} \right)^{3-\frac{\beta^2}{2} K} + ... \nn
\end{eqnarray}
The ellipsis denotes higher order terms in $\frac{da}{a}$. $K_\textit{eff}$ must remain constant with respect to changes in the low energy scale $a \rightarrow  a + da$. The Luttinger parameter $K$ and the coupling $y$ must flow to accomodate these changes:
\begin{equation}
K(a + da) = K(a) - \frac{\beta^2 y^2 K^2}{2^4} \frac{da}{a}.
\end{equation}
The rescaling of the integrand yields the equation for $y$
\begin{equation}
y^2( a + da )  = y^2( a ) \left( \frac{a+da}{a} \right)^{4-\frac{\beta^2}{2}K(a)}.
\end{equation}
Changing variable such that $a(l) = a e^l$ yields the following equations
\begin{eqnarray}
\frac{dK}{dl} = -\frac{\beta^2}{2^4} y^2 K^2, \;\;
\frac{dy}{dl} = \left( 2 -\frac{\beta^2}{4} K \right) y.
\end{eqnarray}
In the weak-coupling limit we approximate $K(l) \approx K(l = 0)$ and the second equation can be integrated to leading order in $y$. To obtain the analogous equations for $\cos (\beta \theta(x))$, one needs to simply map $K \rightarrow K^{-1}$ in all equations.

As the sine-Gordon term flows to strong coupling, the spectrum will acquire a gap $\Delta$, determined as follows. We define the parameter $l^*$ at which $y$ flows to strong coupling:
\begin{equation}
y(l^*) = 1 = \frac{g a}{u} e^{\left(2 - \frac{\beta^2}{4}K \right)l^*}.
\end{equation}  
Then, we use the fact that within our notations the gap is defined as:
\begin{equation}
l^* = \ln \left( \frac{u}{\Delta a}\right)
\end{equation}
The asymptotic form for the gap $\Delta$ then is
\begin{equation}
\Delta \sim \frac{u}{a} y ^ \frac{1}{2 - \frac{\beta^2}{4}K}.
\end{equation}
If the sine-Gordon term was instead $\int dx \cos(\beta \theta)$, then this would be modified by replacing $K \rightarrow K^{-1}$.


\section{Second order Hamiltonian for spinful fermion ladders}
\label{Ap:And}
This appendix contains the derivation of Eqs.~(\ref{Eq:FHSG}) and~(\ref{Eq:FHSG2}). For brevity, we omit from the equations containing Hamiltonians the spatial integrals $\int dx ...$, i.e. all equations in this appendix represent Hamiltonian densities. Using Eq.~(\ref{Eq:FermiFieldAlpha}) we write the hopping term $H_\perp$ in Eq.~(\ref{Eq:HLat}) in the continuum limit as
\begin{widetext}
\begin{eqnarray}
\label{Eq:IntHperp}
\mathcal{H}^{I}_\perp &=& - g a \sum_\sigma \sum_{r,r'} \left[ (\psi^1_{r,\sigma})^\dagger \psi^2_{r',\sigma} e^{ia'A_{\perp \sigma}} + \text{H.c.} \right]
 \\
&=& -\frac{g}{2 \pi} \sum_\sigma   e^{i a' A_{\perp\sigma}} \times  \Big\{ e^{ i \left[ - \phi^-_{\rho} -\theta^-_{\rho} + \sigma \left( - \phi^-_{\sigma} - \theta^-_{\sigma} \right)  \right] } e^{ i ( k^1_{F} - k^2_{F} ) x }  + e^{ i \left[ - \phi^+_{\rho} -\theta^-_{\rho} + \sigma \left( - \phi^+_{\sigma} - \theta^-_{\sigma} \right)  \right] } e^{ i ( k^1_{F} + k^2_{F} ) x } \nonumber \\
&&\;\;\;\;\;\;\;\;\;\;+ e^{ i \left[ + \phi^+_{\rho} -\theta^-_{\rho} + \sigma \left( \phi^+_{\sigma} - \theta^-_{\sigma} \right)  \right] } e^{ -i ( k^1_{F} + k^2_{F} ) x } + e^{ i \left[ + \phi^-_{\rho} -\theta^-_{\rho} + \sigma \left( + \phi^-_{\sigma} - \theta^-_{\sigma} \right)  \right] } e^{ - i ( k^1_{F} - k^2_{F} ) x } \Big\} + \text{H.c.} \nonumber
\end{eqnarray}

Note that this part of the Hamiltonian is irrelevant if either one of the four fields $\phi^{1,2}_\sigma$ or $\phi^{1,2}_\rho$ is gapped (either a spin gap or a charge gap develops in one of the chains).  However, relevant contributions in Eq.~(\ref{Eq:IntHperp}) give rise to non-zero terms at second order in perturbation theory. The effective Hamiltonian density corresponding to order $\frac{g^2}{\Delta_\sigma^1}$ is [see the derivation in Appendix~\ref{Ap:SecondOrderPT}], assuming that a spin gap $\Delta_\sigma^1$ has developed in chain 1:
\begin{eqnarray}
\label{Eq:HIIPT}
&&\mathcal{H}^{II,hf}_\perp = -\frac{\pi a}{\Delta_{\sigma}^1} g^2 
\left\{ \sum_\sigma \left[ ( \psi^1_{-,\sigma})^\dagger \psi_{-,\sigma}^2 + (\psi_{+,\sigma}^1)^\dagger \psi_{+,\sigma}^2   \right] e^{i a' A_{\perp,\sigma} } +\text{H.c.}    \right\}^2  
 \nonumber \\ 
&&\;\;\;\;\;\;\;\;\;\;\;\;\;\;\;\;\;\;\;\;-\frac{\pi a}{\Delta_{\sigma}^1} g^2 \left\{ \sum_\sigma \left[ (\psi^1_{-,\sigma})^\dagger \psi^2_{+,\sigma} + (\psi^1_{+,\sigma})^\dagger \psi^2_{-,\sigma} \right] e^{i a' A_{\perp,\sigma} } + \text{H.c.}   \right\}^2.
\end{eqnarray}
The square of Eq.~(\ref{Eq:IntHperp}) contains as well a set of contributions which are proportional to $\exp( \pm 2 i k_F^\alpha x)$. We have dropped these contributions from Eq.~(\ref{Eq:HIIPT}), since they are oscillatory at or near half-filling in each chain, $k_F^\alpha \approx \frac{\pi}{2a}$. The first term of~(\ref{Eq:HIIPT}) contains terms which are proportional to $\exp\left[ \pm 2 i (k_F^1 - k_F^2 ) x \right]$, whereas the second term contains contributions proportional to $\exp\left[ \pm 2 i (k_F^1 + k_F^2 ) x \right]$. Each of the two terms contains nonoscillatory contributions in addition to the ones mentioned.

From~(\ref{Eq:HIIPT}) we derive the effective Hamiltonian~(\ref{Eq:FHSG}) of Sec.~\ref{Sec:WF}, corresponding to the case $k_F^1 + k_F^2 = \frac{\pi}{a}$, with $k_F^1 \neq k_F^2$ so as to make umklapp terms irrelevant in each chain. In addition, assuming a spin gap in chain 1, $\Delta_\sigma^1$, terms proportional to $\exp(i \theta_\sigma^1)$ are irrelevant. The resulting Hamiltonian at second order in perturbation theory is

\begin{eqnarray}
\mathcal{H}_\perp^{II} = - \frac{g^2}{\Delta_\sigma^1} \frac{1}{\pi a} \Big\{ 
4 &+& 4 \cos( \sqrt{2} \phi_\sigma^1) \cos( \sqrt{2} \phi_\sigma^2 ) \cos( - 2 \theta_\rho^- + a' A_{\perp,\uparrow} + a' A_{\perp,\downarrow}) \nonumber \\
&+& 2 \cos( \sqrt{2} \phi_\sigma^1 + \sqrt{2} \phi_\sigma^2 ) \cos\left[ - 2 \phi_\rho^+ + 2 x(k_F^1 + k_F^2) \right] \nonumber \\
&+& \cos\left[  a' A_{\perp,\downarrow} + a' A_{\perp,\uparrow} + 2 x ( k_F^1 + k_F^2 ) - 2 \theta_\rho^- - 2 \phi_\rho^+  \right] \nonumber \\
&+& \cos\left[  a' A_{\perp,\downarrow} + a' A_{\perp,\uparrow} - 2 x ( k_F^1 + k_F^2 ) - 2 \theta_\rho^- + 2 \phi_\rho^+  \right] \Big\}.
\end{eqnarray}
Note that we have made explicit the oscillatory arguments $2 x ( k_F^1 + k_F^2 )$. At half-filling $k_F^1 + k_F^2 = \frac{\pi}{a}$ these are multiples of $2 \pi$, but we will be concerned in Sec.~\ref{Subsec:FLa2} with slight deviations from half filling. 

From the analogue of~(\ref{Eq:HIIPT}) in the case of a charge gap ($\Delta_\sigma^1 \to \Delta_\rho^1$) we derive the effective Hamiltonian~(\ref{Eq:FHSG2}) of Sec.~\ref{Subsec:FMSp}, corresponding to the case $k_F^1 + k_F^2 = \frac{\pi}{a}$, with $k_F^1 = k_F^2$ so as to make umklapp terms relevant in each chain. In addition, assuming a charge gap $\Delta_\rho^1$, all terms containing $\exp( i \theta_\rho^1)$ are irrelevant. The resulting Hamiltonian at second order in perturbation theory is
\begin{eqnarray}
\mathcal{H}_\perp^{II} = - \frac{g^2}{\Delta_\rho^1} \frac{1}{\pi a} \Big[ 
4 &+& 4 \cos( \sqrt{2} \phi_\sigma^1) \cos( \sqrt{2} \phi_\sigma^2 ) \cos( a' A_{\perp,\downarrow} - a' A_{\perp,\uparrow} + 2 \theta_\sigma^- ) \nonumber \\
&+& 2 \cos( 2 \phi_\sigma^- ) \cos( 2 \phi_\rho^- ) + 2 \cos( 2\phi_\sigma^+ ) \cos( 2 \phi_\rho^+ ) \nonumber \\
&+& 4 \cos( a'A_{\perp,\downarrow} - a'A_{\perp,\uparrow} + 2 \theta_\sigma^- ) \cos( \sqrt{2} \phi_\rho^1) \cos(\sqrt{2} \phi_\rho^2) \Big].
\end{eqnarray}
In Sec.~\ref{Subsec:FMSp} we will assume that $K_\sigma$ is large enough such that the terms favoring the formation of a spin gap are irrelevant, and then the simpler form remains
\begin{equation}
\mathcal{H}_\perp^{II} = -\frac{g^2}{\Delta_\rho^1} \frac{1}{\pi a} \left[  4 + 4 \cos\left(a' A_{\perp,\downarrow} - a' A_{\perp,\uparrow} + 2 \theta_\sigma^-  \right) \cos( \sqrt{2} \phi_\rho^1 ) \cos( \sqrt{2} \phi_\rho^2 ) \right].
\end{equation}
Apart from an additive constant, this is Eq.~(\ref{Eq:FHSG2}) in the main text.

\end{widetext}


\section{Effective Hamiltonian at second order in perturbation theory}
\label{Ap:SecondOrderPT}
In this appendix we derive Eq.~(\ref{Eq:HIIPT}). The general result can be summarized as follows. Let us consider a generic gapped ($\Delta$) unperturbed Hamiltonian and a sine-Gordon perturbation $\mathcal{T}$. Assuming that correlation functions of the perturbation are cut off at separation larger than the correlation length $\xi = u / \Delta$ associated with the gap, an effective sine-Gordon term of order $\mathcal{T}^2 / \Delta$ can approximate expectation values of arbitrary observables at second order in $\mathcal{T}$. 

Let the gapped Hamiltonian density be
\begin{equation}
\mathcal{H}(x) = \mathcal{H}_0[\vartheta(x),\varphi(x)] - \frac{g}{a} \cos( \alpha \vartheta(x) ) + \mathcal{H}_0'[\vartheta'(x),\varphi'(x)]
\end{equation}
where $\mathcal{H}_0(x)$ ($\mathcal{H}_0'(x)$) is the Luttinger liquid Hamiltonian for conjugate fields $\vartheta, \varphi$ ($\vartheta', \varphi'$). Let $\Delta = u / \xi$ be the gap associated with the ordering of  $\vartheta$ due to the cosine potential, where $u$ is the velocity of excitations characterizing $\mathcal{H}_0$. 

Let $\mathcal{T}(x)$ be a sum of sine-Gordon terms added as perturbations to $\mathcal{H}(x)$: 
\begin{equation}
\mathcal{T}(x) = - \frac{t}{a} \left[\cos( \varphi'(x) - \varphi(x) ) + \cos( \varphi'(x) + \varphi(x) ) \right]
\end{equation}
Importantly, the fields that $\mathcal{T}$ would pin do not commute with $\vartheta$ pinned by $\mathcal{H}$, and are therefore irrelevant. However, the term $\cos( 2 \varphi'(x) )$ that appears in $\mathcal{T}^2(x)$, can order $\varphi'$, which commutes with $\vartheta$. At second order in perturbation theory in $\mathcal{T}$ there appear terms which are relevant in the renormalization group sense, although $\mathcal{T}$ itself is irrelevant. 

The addition of $\mathcal{T}$ leads to the effective Hamiltonian density: 
\begin{equation}
\label{Eq:HeffT2}
\mathcal{H}_{\textit{eff}} = \mathcal{H} - \frac{\pi a }{\Delta} \mathcal{T}(x)^2.
\end{equation} 
For any observable $A$, the expectation values with respect to $\mathcal{H} + \mathcal{T}$ and with respect to $\mathcal{H}_\textit{eff}$ are equal up to order $\mathcal{T}^2$
\begin{equation}
\langle A \rangle_{\mathcal{H} + \mathcal{T}} = \langle A \rangle_{\mathcal{H}_\textit{eff}}.
\end{equation} 
In the remainder of this section we argue that this follows from an expansion of the partition function.

Let us denote $\int \mathcal{T} \equiv \int dx d\tau \mathcal{T}(x,\tau)$. The proof of Eq.~(\ref{Eq:HeffT2}) follows from expanding an arbitrary expectation value to second order in $\mathcal{T}$
\begin{widetext}
\begin{eqnarray}
\langle A \rangle_{\mathcal{H} + \mathcal{T}} &=& \frac{\text{Tr}\, \left\{ e^{ - \int dx d\tau (\mathcal{H} + \mathcal{T})} A \right\} }{ \text{Tr}\, \left\{ e^{ - \int dx d\tau (\mathcal{H} + \mathcal{T})} \right\} } \nonumber \\
&=& \left[ \langle A  \rangle_\mathcal{H}  -  \langle (\int \mathcal{T}) A  \rangle_\mathcal{H} + \frac{1}{2}  \langle ( \int \mathcal{T} )^2 A \rangle_\mathcal{H} \right] \, \left[ 1 + \langle \int \mathcal{T} \rangle_\mathcal{H} - \frac{1}{2} \langle (\int \mathcal{T})^2 \rangle_\mathcal{H} + \langle \int \mathcal{T} \rangle_\mathcal{H}^2 \right] + O\left[ (\int \mathcal{T})^3 \right] \nonumber \\
&=& \langle A \rangle_\mathcal{H} + \frac{1}{2} \left[ \langle (\int \mathcal{T})^2 A \rangle_\mathcal{H} - \langle (\int \mathcal{T})^2 \rangle_\mathcal{H} \langle A \rangle_\mathcal{H}  \right] + O\left[ (\int \mathcal{T})^3 \right].
\label{Eq:EVAsTs}
\end{eqnarray}
\end{widetext}
In the last equality we have dropped all contributions at first order in $\mathcal{T}$. These contain the disordered field $\varphi$. 

Now, let 
\begin{equation}
d_{12} = \sqrt{(x_1-x_2)^2 + u^2 (\tau_1 - \tau_2)^2}
\end{equation}
and let $\mathcal{T}_{1} \equiv \mathcal{T}(x_1,\tau_1)$. Consider the following $\mathcal{T} \mathcal{T}$ correlation functions
\begin{eqnarray}
\label{Eq:EVTT}
\langle (\int \mathcal{T})^2 \rangle = \int dx_1 d\tau_1 dx_2 d\tau_2 \langle \mathcal{T}_1 \mathcal{T}_2 \rangle_\mathcal{H}  \nonumber \\=  \int  dx_1 d\tau_1 dx_2 d\tau_2 R( d_{12} ) \exp( -d_{12} / \xi ),
\end{eqnarray}
where $R( d_{12} )$ is a power law and $\xi$ is the correlation length associated with the gapped mode $\vartheta$. Switching to relative and center of mass coordinates, and assuming $\beta \gg 0$, we obtain the approximate form
\begin{eqnarray}
\label{Eq:EVTTAp}
\text{Eq.~(\ref{Eq:EVTT})}= 2\pi \int dX d\eta \int d(d_{12}) \, d_{12}  R( d_{12} ) \exp( -d_{12} / \xi ) \nonumber \\ \;
\approx 2 \pi (\xi - a) R( a ) \int dX d\eta \, 1 \approx 2 \pi \xi R(a) \, (L\beta) \nonumber
\end{eqnarray}
Then we may approximate
\begin{eqnarray}
(\int \mathcal{T})^2 &=& \int dx_1 d\tau_1 dx_2 d\tau_2 \mathcal{T}_1 \mathcal{T}_2  \nonumber \\ 
&\approx& \frac{2\pi \xi a}{u} \int dX d\eta \, \mathcal{T}(X, \eta) \mathcal{T}(X+a,\eta) \nonumber \\ &=& \frac{2\pi a}{\Delta} \int dX d\eta \, \mathcal{T}(X, \eta) \mathcal{T}(X+a,\eta) \nonumber \\
&=& \frac{2\pi a}{\Delta} \int \mathcal{T}^2
\end{eqnarray}
By replacing $(\int \mathcal{T})^2 \approx \frac{2\pi a}{\Delta} \int \mathcal{T}^2$ in Eq.~(\ref{Eq:EVAsTs}), we obtain the lowest order contribution, of order $\mathcal{T}^2$, to the expectation value computed with respect to $\mathcal{H}_\textit{eff}$. This proves Eq.~(\ref{Eq:HeffT2}).


\section{Effective edge Hamiltonian}
\label{Ap:EdgeTheory}
In this Appendix, we derive the effective edge theory summarized in Sec.~\ref{Subsubsec:EdgeTheory}. The Hamiltonian is expressed in Eq.~(\ref{Eq:FullHLaughlinCoupling}), which we reproduce here
\begin{eqnarray}
\label{ApEq:H}
\mathcal{H} = \frac{v}{2\pi} \sum_\alpha \int dx \left[ K (\nabla \theta^\alpha)^2 + \frac{1}{K} (\nabla \phi^\alpha)^2  \right] \nonumber \\ 
  + \frac{V_\perp a}{\pi^2} \int dx (\nabla \phi^1) (\nabla \phi^2)  \nonumber \\
 - 2 g n_0 \int dx \cos( \theta^1 - \theta^2  - m \phi^1 - m \phi^2). 
\end{eqnarray}
To obtain the effective gapless Hamiltonian describing the edge chiral fields $\phi_{-1} \equiv \phi_{-1}^2$ and $\phi_{+1} \equiv \phi_{+1}^1$, we integrate out the gapped degrees of freedom in~(\ref{ApEq:H}). Importantly, this relies on the assumption that the coupling constant $g$ is relevant, which assumes strong enough $V_\perp$ (or long ranged intrachain repulsion), as explained in the main text.

Readers may wish to skip the detailed calculation and go directly to Eq.~(\ref{ApEq:Hedge}), for the effective Hamiltonian $\mathcal{H}_\textit{edge}$, and the discussion thereafter.

We recall here the linear transformations of fields in Eqs.~(\ref{Eq:ChiralFields}) and (\ref{Eq:BulkFields}):
\begin{eqnarray}
\phi_{r}^\alpha &=& \theta^\alpha/m + r \phi^\alpha, \alpha = 1,2, r = \pm 1, \nonumber \\
\phi &=& (-\phi^1_{-1} + \phi^2_{+1})/2, \theta = (\phi^1_{-1} + \phi^2_{+1})/2.
\end{eqnarray}
The bulk fields obey the following algebra:
\begin{equation}
[ \phi(x), \theta(x')] = i \frac{\pi}{2m} \text{Sign}(x'-x), 
\end{equation}
therefore the momentum associated to $\phi$ is $\Pi_\phi = \frac{m}{\pi} \nabla \theta$. The chiral fields obey:
\begin{equation}
[\phi^\alpha_r(x), \phi^\beta_p(x') ] = i r \delta_{\alpha\beta} \delta_{rp} \frac{\pi}{m} \text{Sign}(x'-x).
\end{equation}
The inverse transformations are
\begin{eqnarray}
\theta^\alpha &=& \frac{m}{2} \left( \phi_{+1}^\alpha + \phi_{-1}^\alpha  \right),  \nonumber \\
\phi^\alpha &=& \frac{1}{2} \left( \phi_{+1}^\alpha - \phi_{-1}^\alpha \right), \nonumber \\
\phi^2_{+1} &=& \phi + \theta,  \nonumber \\ 
\phi^1_{-1} &=& - \phi + \theta. 
\end{eqnarray}
for $\alpha = 1,2$. 

In terms of these variables, the sine-Gordon term is $- 2 g n_0 \int dx \cos( 2 m \phi)$. Let us assume that $g$ is a large energy scale, so that we can approximate the sine Gordon term by the quadratic contribution, i.e.
\begin{equation}
- 2 g n_0 \int dx \cos( 2 m \phi) \approx 4 m^2 g n_0 \phi^2 + \text{const.}
\end{equation}

In terms of $\phi,\theta, \phi_{+1} \equiv \phi^1_{+1}, \phi_{-1} \equiv \phi^{2}_{-1}$, the various contributions to Eq.~(\ref{ApEq:H}) are
\begin{widetext}
\begin{eqnarray}
\sum_\alpha (\nabla \theta^\alpha)^2 &=& \frac{m^2}{2} \left[ (\nabla \phi)^2 + (\nabla \theta)^2 \right] + \frac{m^2}{2} \left[ (-\nabla \phi_{+1} + \nabla \phi_{-1}) (\nabla \phi)  + (\nabla \phi_{+1} + \nabla \phi_{-1}) (\nabla \theta)  \right]  + \frac{m^2}{4} \left[ (\nabla \phi_{-1})^2 + (\nabla \phi_{+1})^2 \right], \nonumber \\
\sum_\alpha (\nabla \phi^\alpha)^2 &=& \frac{1}{2} \left[ (\nabla \phi)^2 + (\nabla \theta)^2 \right] - \frac{1}{2} \left[ (-\nabla \phi_{+1} + \nabla \phi_{-1}) (\nabla \phi)  + (\nabla \phi_{+1} + \nabla \phi_{-1}) (\nabla \theta)  \right]  + \frac{1}{4} \left[ (\nabla \phi_{-1})^2 + (\nabla \phi_{+1})^2 \right], \nonumber \\
(\nabla \phi^1 )( \nabla \phi^2 ) &=& \frac{1}{4} ( \nabla \phi_{+1} + \nabla \phi )( \nabla \phi - \nabla \phi_{-1}) + \frac{1}{4} \nabla \theta (  \nabla \phi_{-1} + \nabla \phi_{+1} ) - \frac{1}{4} ( \nabla \theta )^2.
\end{eqnarray}
\end{widetext}

To obtain the effective dynamics $\phi_{-1}$ and $\phi_{+1}$ only, we integrate out $\theta$, then the gapped mode $\phi$. This integration is easily performed in Fourier space. We use the Fourier transform convention
\begin{eqnarray}
f(r) = \frac{1}{\beta L} \sum_q f_q e^{i q r},
\end{eqnarray}
where $r=(x,v\tau)$, $q=(k,\omega_n/v)$, $qr=kx-\omega_n \tau$, and the Matsubara frequencies are $\omega_n = \frac{2\pi n}{\beta}$ for all integer $n$. The part of the action that depends on $\theta$ is
\begin{equation}
\beta L S_\theta = \sum_q \frac{i m k \omega_n}{\pi} \phi_k \theta^*_k + \sum_q k^2 \theta_k F^*_k + \sum_q G_k \theta_k^* \theta_k.
\end{equation}
We defined 
\begin{eqnarray}
F_k &=& \left[ \frac{V_\perp a}{4 \pi^2} + \frac{v}{4\pi} \left( K m^2 - \frac{1}{K} \right)  \right] ( \phi_{+1,k} + \phi_{-1,k})  \nonumber \\
G_k &=& -\frac{V_\perp a}{4 \pi^2} + \frac{v}{4\pi} \left( K m^2 + \frac{1}{K} \right).
\end{eqnarray}
For any field $f$ that is a real-valued function of the coordinate $x$, the Fourier transform has the property $f_q = f_{-q}^*$. The $\theta$ integral is gaussian. We remark that the $\phi$ integral introduces terms of order $1/g$. We make the assumption that $g$ is large, and find that all contributions from the $\phi$ integral contain terms quartic in derivatives, which we drop. Fourier transforming back to real space, the remaining effective action yields the following Hamiltonian for the ``edge'' degrees of freedom
\begin{equation}
\label{ApEq:Hedge}
\mathcal{H}_\textit{edge} = \int dx \left\{  A \left[ (\nabla \phi_{+1})^2 + (\nabla \phi_{-1})^2 \right] + B (\nabla \phi_{+1})(\nabla \phi_{-1})   \right\}
\end{equation}
where the coefficients are given by
\\
\begin{eqnarray}
A &=& \frac{v}{8 \pi } \left( K m^2 + \frac{1}{K} \right) + \bar{A} \nonumber \\
B &=& \frac{V_\perp a}{2\pi^2} + \bar{A} \nonumber \\
\bar{A} &=& \frac{1}{4\pi} \frac{\left[ \frac{V_\perp a}{\pi} + v (K m^2 - 1/K) \right]^2 }{-\frac{V_\perp a}{\pi} + v (K m^2 + 1/K)}.
\end{eqnarray}
Equivalently, these describe a Luttinger liquid with effective velocity of excitations and Luttinger parameter 
\begin{equation}
v_{\textit{eff}}^2 = (A - B)( A + B),\; K_{\textit{eff}}^2 = ( A - B ) / ( A + B ).
\end{equation}

It is instructive to consider a simple case. In the limit $K = 1/m$, there must be no backscattering term in chain $\alpha$ of the form $(\nabla \phi^\alpha_{+1})(\nabla \phi^\alpha_{-1})$. Note as well that this value corresponds to long ranged repulsive interactions in each of the bosonic chains \cite{Giamarchi}. Let us assume that $V_\perp$ is small while the sine-Gordon term in~(\ref{ApEq:H}) is relevant in the renormalization group flow. For small enough $V_\perp$, we approximate 
\begin{eqnarray}
A \approx  \frac{m v}{4 \pi}, \; B \approx \frac{V_\perp a}{2\pi^2}, \; \bar{A} = O(V_\perp^2).
\end{eqnarray}
That is, all backscattering terms in the effective Hamiltonian arise from $V_\perp$. The resulting Luttinger liquid Hamiltonian is characterized by the following velocity and Luttinger parameter
\begin{eqnarray}
v_{\textit{edge}}^2 &=& (\pi mv - 2 V_\perp a) (\pi mv + 2 V_\perp a), \nonumber \\ 
K_{\textit{edge}}^2 &=& \frac{\pi mv - 2 V_\perp a}{\pi mv + 2 V_\perp a}. 
\end{eqnarray}
A useful check is to set $V_\perp=0$. In this limit all backscattering terms between the remaining gapless chiral fields must be absent, and the edge is described by the chiral Luttinger liquid Hamiltonian,
\begin{equation}
\mathcal{H}_\textit{edge} = \frac{m v}{4\pi} \int dx \left[ (\nabla \phi_{+1})^2 + (\nabla \phi_{-1})^2   \right].
\end{equation}

\bibliography{references}

\end{document}